\documentclass[
  final,
  onefignum,
  onetabnum,
]{siamart250211}

\usepackage[utf8x]{inputenc}

\usepackage{lipsum}
\usepackage{amsfonts}
\usepackage{amssymb}
\usepackage{graphicx}
\usepackage{algorithmic}
\ifpdf
  \DeclareGraphicsExtensions{.eps,.pdf,.png,.jpg}
\else
  \DeclareGraphicsExtensions{.eps}
\fi
\usepackage{siunitx}
\usepackage{csquotes}
\usepackage{microtype}
\usepackage{booktabs}

\newcommand*{\vb}[1]{\boldsymbol{#1}}  % vectors
\newcommand*{\mat}[1]{\vb{\mathsf{#1}}}

\newcommand*{\VortexRadius}{a}

\newcommand*{\CoreSizeCutoffCoef}{\epsilon}
\newcommand*{\err}{\varepsilon}
\newcommand*{\abserr}{\err}
\newcommand*{\relerr}{\err^{\text{rel}}}
\newcommand*{\tol}{\err_{\text{tol}}}

\newcommand*{\im}{\mathrm{i}}  % imaginary unit

\newcommand*{\Xvec}{\vb{X}}
\newcommand*{\cvec}{\vb{c}}
\newcommand*{\kvec}{\vb{k}}
\newcommand*{\nvec}{\vb{n}}
\newcommand*{\pvec}{\vb{p}}
\newcommand*{\qvec}{\vb{q}}
\newcommand*{\uvec}{\vb{u}}
\newcommand*{\rvec}{\vb{r}}
\newcommand*{\svec}{\vb{s}}
\newcommand*{\vvec}{\vb{v}}
\newcommand*{\xvec}{\vb{x}}
\newcommand*{\yvec}{\vb{y}}
\newcommand*{\vortvec}{\vb{\omega}}
\newcommand*{\zerovec}{\vb{0}}

\newcommand*{\dmin}{\ell_{\text{min}}}
\newcommand*{\Tkw}{T_{\text{kw}}}

\newcommand*{\psivec}{\vb{\psi}}

\newcommand*{\dd}{\mathrm{d}}
\newcommand*{\deriv}[2]{\frac{\dd #1}{\dd #2}}
\newcommand*{\gradient}{\vb{\nabla}}
\newcommand*{\laplacian}{\nabla^2}
\newcommand*{\curl}{\gradient \times}

\newcommand*{\local}{_{\text{local}}}

\newcommand*{\self}{_{\text{self}}}
\newcommand*{\reference}{^{\text{ref}}}
\newcommand*{\fit}{_{\text{fit}}}

\newcommand*{\shortrangeText}{\text{(n)}}
\newcommand*{\longrangeText}{\text{(f)}}
\newcommand*{\shortrange}{^{\shortrangeText}}
\newcommand*{\longrange}{^{\longrangeText}}
\newcommand*{\sr}{\shortrange}
\newcommand*{\lr}{\longrange}
\newcommand*{\rcut}{r_{\text{c}}}
\newcommand*{\kmax}{k_{\text{max}}}

\DeclareMathOperator{\erf}{erf}
\DeclareMathOperator{\erfc}{erfc}

\newcommand*{\Ccal}{\mathcal{C}}
\newcommand*{\VortexLength}{\mathcal{L}}
\newcommand*{\tparam}{\tau}
\newcommand*{\Tparam}{\mathcal{T}}

\newcommand*{\Nfil}{N_{\text{f}}}  % number of filaments

\newcommand*{\NumberOfEllipses}{40}

\newsiamremark{remark}{Remark}
\newsiamremark{hypothesis}{Hypothesis}
\crefname{hypothesis}{Hypothesis}{Hypotheses}
\newsiamthm{claim}{Claim}

\newcommand{\fulltitle}{Fast and accurate evaluation of Biot--Savart integrals over spatial curves in periodic domains}
\newcommand{\shorttitle}{Fast and accurate evaluation of Biot--Savart integrals}
\headers{\shorttitle}{J. I. Polanco}

\title{\fulltitle}

\author{%
  Juan Ignacio Polanco\thanks{Univ.\ Grenoble Alpes, CNRS, Grenoble INP, LEGI, 38000 Grenoble, France
  (\email{juan-ignacio.polanco@cnrs.fr}).}
}

\usepackage{amsopn}

\ifpdf%
\hypersetup{
  pdftitle={\fulltitle},
  pdfauthor={J. I. Polanco}
}
\fi

\begin{document}

\maketitle

% REQUIRED
\begin{abstract}
  The Biot--Savart law is relevant in physical contexts including
  electromagnetism and fluid dynamics.
  In the latter case, when the rotation of a fluid is confined to a set of very thin
  vortex filaments, this law describes the velocity field induced by the
  spatial arrangement of these objects.
  The Biot--Savart law is at the core of vortex methods used in the simulation
  of classical and quantum fluid flows.
  Naïve methods
  are inefficient when dealing with large numbers of vortex elements, which makes them
  inadequate for simulating turbulent vortex flows.
  Here we exploit a direct analogy between the Biot--Savart law and
  electrostatics to adapt Ewald summation methods, routinely used in molecular
  dynamics simulations, to vortex filament simulations in three-dimensional periodic domains.
  In this context, the basic idea is to split the induced velocity onto
  (i) a coarse-grained velocity generated by a Gaussian-filtered vorticity field, and
  (ii) a short-range correction accounting for near-singular behaviour near the vortices.
  The former component can be accurately and efficiently evaluated using the
  nonuniform fast Fourier transform algorithm.
  Analytical accuracy estimates are provided as a function of the
  parameters entering the method.
  We also discuss how to properly account for the finite vortex core size in kinetic energy estimations.
  Using numerical experiments, we verify the accuracy and the conservation
  properties of the proposed approach.
  Moreover, we demonstrate the $O(N \log N)$ complexity of the method
  over a wide range of problem sizes $N$, considerably better than the $O(N^2)$ cost of a naïve approach.
\end{abstract}

% REQUIRED

% REQUIRED

\section{Introduction}

The Biot--Savart law is well known for describing the magnetic field generated by
a steady electric current.
It is also very relevant in fluid dynamics, where it allows to obtain the fluid
velocity induced by a known vorticity field.
We are interested here in the specific case where the electric current or the
vorticity field are confined to spatial curves in three-dimensional space.
This is a commonly encountered problem in undergraduate electromagnetism
lectures, where the question is what is the magnetic field induced by a current
flowing through a thin conducting wire.
In the fluid dynamics case, the equivalent would be a very thin vortex
filament, which can be a reasonable idealised model for vortices found in
the small-viscosity limit.
In fact, it is also a very accurate description of vortices in certain superfluids such
as liquid helium-4 near absolute zero, where rotational motion is confined
to so-called quantum vortices of atomic-size thickness (the vortex core radius
is $\VortexRadius \approx \SI{e-10}{\metre}$), and the circulation (or strength) of each
vortex takes a constant value $\kappa \approx \SI{9.97e-8}{m^2/s}$ dictated by
quantum mechanical constraints.
The velocity induced at a point $\xvec$ away from a vortex core is then given
by the Biot--Savart law,
\begin{equation}
  \vvec(\xvec) =
  \frac{\kappa}{4\pi}
  \oint_{\Ccal} \frac{(\svec - \xvec) \times \dd \svec}{|\svec - \xvec|^3},
  \label{eq:BS_intro}
\end{equation}
where $\Ccal$ denotes one or more oriented curves representing the vortex
filament geometry, and $\svec \in \Ccal$ denotes a vortex location.

In classical fluid dynamics, the Biot--Savart law is at the core of vortex
methods~\cite{Cottet2000, Koumoutsakos2005} used to describe incompressible
viscous flows at relatively high Reynolds numbers.
These are commonly used in aerodynamics applications~\cite{Vermeer2003}, but
have also been applied to problems as varied as the simulation of
self-propelled swimmers in viscous fluids~\cite{Gazzola2014}.
Note that, in viscous fluids, vortex methods need to account for effects
including vorticity diffusion and energy dissipation due to viscosity.
To achieve this, these methods typically deal with vortex \emph{particles},
which should not be interpreted as physical objects but as an ensemble of
point charges generating a fluid flow.
In particular, the connectivity of vortex particles is not relevant to vortex
particle methods~\cite{Koumoutsakos2005}.

Here we focus on the conceptually simpler application of vortex methods to
superfluid helium-4, where viscous effects are absent near absolute zero
and vortex filaments are in fact the main physical object of interest.
In this context, the Biot--Savart law is the basic ingredient of the vortex filament model
(VFM), which is one of the most common approaches for describing superfluid
flows~\cite{Schwarz1985, Hanninen2014, Barenghi2023}.
This model is valid at scales much larger than the atomic vortex thickness, and
is therefore well adapted for describing macroscopic vortex motion.
Numerically, the standard approach for representing vortex filaments consists
in discretising them as a series of connected vortex points.
The connectivity is required in order to obtain derived geometrical information
such as local tangents and curvatures.
Each vortex point evolves in time according to the Biot--Savart law.
In fact, the integral \cref{eq:BS_intro} diverges when evaluated on a vortex
point $\xvec = \svec_0$.
The singularity can be avoided by taking into
account the finite (but small) vortex core radius
$\VortexRadius$~\cite{Schwarz1985},
which introduces a logarithmic dependence of the velocity on $a$ (see \cref{sec:VFM}).
It is worth mentioning that this model shares many similarities with slender
body theory~\cite{Batchelor1970, Lauga2009, Mori2020}, which describes
the dynamics of thin fibers in Stokes flows.
Similarly to the VFM, the forces induced by a 3D fiber on the fluid are
approximated by a distribution of singularities along the fiber centreline.
Furthermore, the velocity induced by a fiber on itself also includes a logarithmic
dependence on its small thickness $a$.

This work is motivated by the study of quantum turbulence~\cite{Barenghi2023},
which is a state of superfluid flows characterised by a wide range of
energetically active length scales.
Numerically, to investigate such a turbulent state, one needs to compute the
non-local interactions between large numbers of vortices, requiring in
particular a large number $N$ of discrete vortex points.
If one explicitly accounts for all pair interactions,
obtaining the velocities at the $N$ points has a $O(N^2)$ cost, which quickly
limits the size of the systems which can be numerically studied.
In the context of particle simulations, this problem has been solved for a long
time, using techniques such as Barnes--Hut (BH) trees~\cite{Barnes1986} or the fast
multipole method (FMM)~\cite{Greengard1987}, which reduce the complexity to $O(N \log N)$
and $O(N)$ respectively.
Such techniques have also been applied to vortex methods for classical~\cite{Cottet2000} and quantum~\cite{Baggaley2012f}
fluids.

Here we are interested in periodic infinite systems, in which
a finite set of $\Nfil$ vortices is replicated an infinite number of times in a
spatially periodic fashion.
Periodic boundary conditions are commonly used to model a variety of physical
systems when one wants to describe phenomena far from boundaries.
A natural way of dealing with periodicity in Cartesian domains is via a
Fourier series representation, as done for example in Fourier pseudo-spectral
methods~\cite{Canuto1988, Boyd2001}.
These take advantage of the fast Fourier transform (FFT) to
efficiently evaluate non-local operations.
This suggests the idea of using FFTs to evaluate costly far-field
interactions between particles or vortices in periodic systems.
However, since particles and vortices are respectively represented as 0D and 1D singularities,
one cannot directly describe the associated source fields (e.g.\ electric
charge density or vorticity) using a truncated Fourier series, which in practice precludes
the use of FFTs.
In particle simulations, one way around this issue is provided by \emph{fast Ewald
summation} methods, which are commonly used in molecular dynamics simulations to speed-up
the evaluation of electrostatic interactions between charged particles in
periodic systems~\cite{Ewald1921, Hockney1988, Darden1993, Deserno1998,
Arnold2005}.
There, the basic strategy is to additively split the singular source field onto a smooth
field responsible for far-field interactions and a correction field accounting
for interactions between nearby particles.
Fast Ewald summation methods are characterised by a $O(N \log N)$ complexity in
the number of particles $N$.
In fluid dynamics contexts, fast Ewald methods have been used to describe 2D
point vortex dynamics~\cite{Monaghan1993}, fluid motions induced by
localised forces in the Stokes regime~\cite{Saintillan2005, Lindbo2010, Maxian2021}, and Darcy
flows at the interface between two immiscible fluids~\cite{Ambrose2013}.
In molecular dynamics simulation benchmarks~\cite{Arnold2013}, it has been
observed that FFT-based Ewald methods can perform slightly better than the
periodic FMM at the same accuracy, actually displaying near linear complexity over a
wide range of problem sizes $N$.
Moreover, when running at low accuracy levels, the
latter can display a slight energy drift over time which is not observed in the
former~\cite{Arnold2013}.

The aim of this paper is to adapt Ewald methods to the evaluation of the
Biot--Savart integral \cref{eq:BS_intro} in three-dimensional periodic systems
and evaluate the relevance of this approach.
The paper begins in \cref{sec:VFM} with an introduction to the VFM in the context of
quantum vortex dynamics.
We also propose an approach for accurately estimating the kinetic energy in
periodic systems.
\Cref{sec:ewald_bs} describes the Ewald-based method used to evaluate
Biot--Savart integrals.
Our approach takes advantage of the nonuniform fast Fourier transform (NUFFT)
algorithm to speed-up computations.
In \cref{sec:truncation_errors}, we provide analytical estimates of the
approximation errors incurred by the proposed approach in terms of tunable
parameters.
The relevance of these estimates is then verified in
\cref{sec:numerical_experiments} using numerical experiments of different test
cases.
That section also showcases the accuracy of energy and impulse conservation by
the method, and finishes with numerical evidence of near linear $O(N \log N)$ complexity
over a wide range of problem sizes $N$.
Finally, \cref{sec:conclusions} is devoted to conclusions.

\section{The vortex filament model}\label{sec:VFM}

The VFM introduced by Schwarz~\cite{Schwarz1985} is one
of the main approaches used to describe theoretically and numerically the
three-dimensional hydrodynamics of quantum fluids such as low-temperature
liquid helium~\cite{Hanninen2014, Barenghi2023}.
Unlike classical fluids, quantum fluids near absolute zero are
characterised by having zero viscosity and being irrotational almost
everywhere.
In fact, rotational motions are confined to very thin vortex filaments carrying
a quantised circulation $\kappa = h / m$, where $h$ is Planck's constant and
$m$ the mass of one atom in the case of helium-4.
In other words, a straight vortex filament induces a circular motion of the fluid around it with
velocity $v = \kappa / (2\pi r)$, where $r$ is the distance to the vortex.

We restrict our attention to the zero temperature limit.
Indeed, in finite temperature superfluid helium, quantum vortices can be seen as
coexisting and interacting with a viscous normal fluid~\cite{Barenghi2014a}.
This is described by more complex models (some of them based on the VFM) which are
still the subject of active research~\cite{Barenghi2023}.
A second very important aspect which is not discussed here is the
reconnection of vortices when they are close to collision.
While this phenomenon is crucial for describing energy dissipation and quantum turbulence,
its modelling is orthogonal to the subject of this work, and can be disregarded
as long as vortex elements stay sufficiently far from each other.

\subsection{Filaments as oriented curves}%
\label{sec:filaments_as_oriented_curves}

In the VFM, one is interested in describing the motion induced by a collection
of vortex filaments on themselves.
Each filament is represented as an oriented curve $\left\{
\svec(\xi), 0 \le \xi \le \VortexLength \right\}$ where $\svec(\xi)$ is a
location on the curve, $\xi$ is the arc length, and $\VortexLength$ is the
total length of the curve.
Here the curve is represented using the natural arc length parametrisation, such that the
unit tangent to the curve is $\svec'(\xi) = \dd\svec / \dd\xi$ with
$|\svec'(\xi)| = 1$ for all $\xi$.
In practice, it is often more convenient to deal with arbitrary parametrisations
which will be denoted $\svec(\tparam)$ for $0 \le \tparam \le \Tparam$, such that
$|\dd\svec / \dd\tparam| \ne 1$ in general.

We will consider the vortex filaments to be embedded in a
triply periodic domain, so that every filament is replicated an infinite number of
times along each Cartesian direction.
For simplicity, throughout this paper, the domain period is set to $L$ in all
directions, but all definitions and results can be readily generalised to
different periodicities in each direction.
In fluid dynamics, Helmholtz' theorems~\cite{Helmholtz1858} state that vortex lines cannot end in
the fluid.
Therefore, in the absence of solid boundaries, 
they must either be closed curves or extend to infinity.
Hence we only consider these two cases, with the additional restriction
that infinite curves must be described by a periodic function matching the domain periodicity.
Both cases are defined by the property
$\svec(\Tparam + \tparam) = \svec(\tparam) + \nvec L$ with $\nvec \in \mathbb{Z}^3$.
In particular, closed curves satisfy this property with $\nvec = \zerovec$.
An example of an infinite periodic curve is $\svec(\tparam) = [\cos(2\pi \tparam / L), 0, \tparam]$
with $\Tparam = L$.
This curve satisfies the above property with $\nvec = [0, 0, 1]$, extending
infinitely along the third Cartesian direction.

\subsection{The Biot--Savart law}
\label{sec:bs_periodic}

We now consider a set of $\Nfil$ vortex filaments and their periodic images,
and denote $\Ccal = \cup_{j = 1}^{\Nfil} \Ccal_j$ the set of spatial curves
$\Ccal_j$ representing their locations.
The vorticity field $\vortvec = \curl \vvec$ associated to these
filaments is then
\begin{equation}
  \vortvec(\xvec) =
  \kappa \sum_{\nvec \in \mathbb{Z}^3}
  \oint_{\Ccal} \delta(\svec - \xvec + \nvec L) \, \dd \svec
  \label{eq:vfm_vorticity}
\end{equation}
where $\delta(\xvec)$ is the Dirac delta function, and the infinite sum over
$\nvec$ accounts for the periodic vortex images.
Here $\kappa$ is the circulation of the vortex filaments, related to the
magnitude of the velocity induced by them.
Throughout this work $\kappa$ will be taken to be constant (as is actually the
case of quantum vortices), but in principle one could also consider a variable
circulation $\kappa(\svec)$.

Inverting the curl operator leads to the Biot--Savart law \cref{eq:BS_intro}
describing the velocity induced by the set of vortex filaments on a point
$\xvec$.
Including periodicity and finite core size effects, this law writes
\begin{equation}
  \vvec(\xvec) =
  \frac{\kappa}{4\pi} \sum_{\nvec \in \mathbb{Z}^3}
  \oint_{\Ccal} \frac{(\svec - \xvec + \nvec L) \times \dd \svec}{|\svec - \xvec + \nvec L|^3} \, \varphi_v(\svec - \xvec + \nvec L).
  \label{eq:BS}
\end{equation}
Here $0 \le \varphi_v(\rvec) \le 1$ accounts for the finite radius
$\VortexRadius$ of the vortex core.
As detailed in \cref{sec:desingularisation_velocity}, it allows to
desingularise the Biot--Savart integral when evaluating the velocity on a vortex.
For all practical purposes, it is constant and equal to $1$ whenever $\xvec$ is not located on a vortex.

Note that periodicity requires the total vorticity to be zero within a periodic
cell, or otherwise the curl operator cannot be inverted (the velocity diverges).
This condition corresponds to $\oint_{\Ccal} \dd\svec = \zerovec$.
This is trivially satisfied for closed filaments, while it requires special care
when dealing with infinite unclosed filaments.

\subsection{Desingularisation of the Biot--Savart integral}%
\label{sec:desingularisation_velocity}

In the VFM, one is generally interested in the velocity induced by the vortex
filaments \emph{on the filaments themselves}.
In other words, one wants to evaluate \cref{eq:BS} on positions
$\xvec = \svec_0 \in \Ccal$, where the Biot--Savart integral clearly
diverges~\cite{Callegari1978}.
More precisely, performing a Taylor expansion of $\svec(\xi)$,
one can show that close to $\svec_0 = \svec(\xi_0)$ the integrand behaves
as~\cite{Arms1965}
\begin{equation}
  \frac{[\svec(\xi) - \svec_0] \times \svec'(\xi)}{
    |\svec(\xi) - \svec_0|^3
  } \, \dd\xi
  =
  \frac{\svec_0' \times \svec_0''}{2 |\xi - \xi_0|} \, \dd\xi + O(1),
  \label{eq:bs_integrand_taylor}
\end{equation}
where primes denote derivatives with respect to the arc length $\xi$.
In particular, $\svec_0' \equiv \svec'(\xi_0)$ and $\svec_0'' \equiv \svec''(\xi_0)$ are
respectively the unit tangent and the curvature vectors at $\svec_0$.

The divergence of the Biot--Savart integral is unphysical since
\cref{eq:vfm_vorticity} does not account for the finite (but very small) radius
$\VortexRadius$ of the vortex core.
This issue can be circumvented in a physically consistent manner by defining
the $\varphi_v$ term appearing in \cref{eq:BS} such that
$\varphi_v(\rvec) = 0$ for $|\rvec| < \CoreSizeCutoffCoef_v \VortexRadius$, and
$1$ otherwise,
where $\CoreSizeCutoffCoef_v$ is a cut-off coefficient.
In other words, $\varphi_v$ is defined such that locations $\svec$ close
to the singularity at $\svec_0$ are omitted from the
integral\footnote{%
  Interestingly, in the context of Stokes flows, slender body theory has been
  recently reformulated using a similar cut-off procedure~\cite{Maxian2021}.
}~\cite[Ch.\,11]{Saffman1993}.
It is convenient to write $\CoreSizeCutoffCoef_v = e^{\Delta}/2$,
where $\Delta$ is a constant which depends on the actual vorticity profile
within the vortex core.
This constant can be analytically derived for commonly used vorticity profiles,
under the assumption that the local curvature radius $1 / |\svec''|$ stays much
larger than the core size $\VortexRadius$.
In particular, $\Delta = 1/4$ for a circular vortex core with uniform vorticity~\cite[Ch.\,10]{Saffman1993}.

\subsection{Discretisation of the Biot--Savart integral}

\begin{figure}
  \begin{center}
    \includegraphics[width=0.5\textwidth]{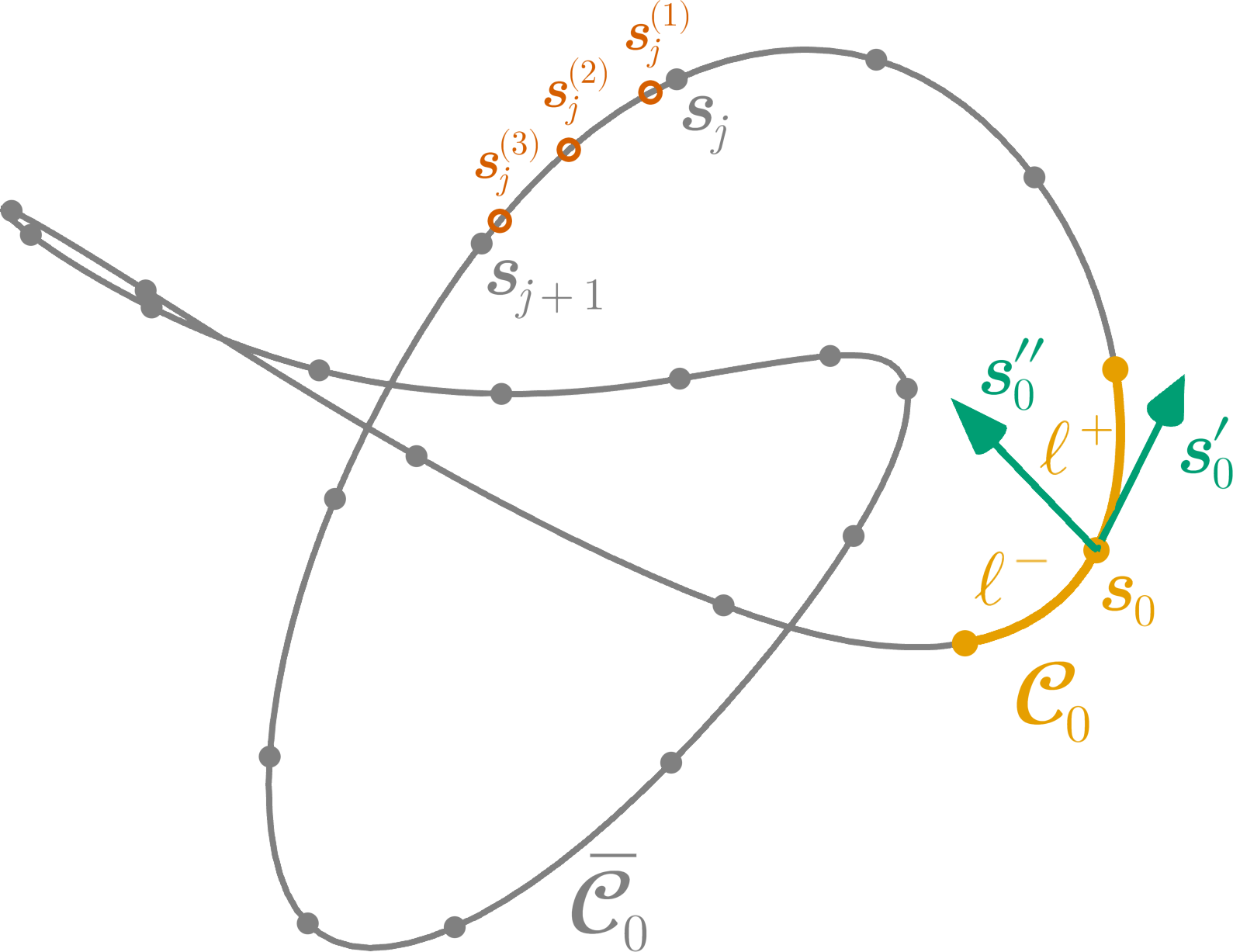}
  \end{center}
  \caption{
    Discretisation of a closed vortex filament, and distinction between local and
    non-local contributions to the Biot--Savart law.
    Filled grey circles represent vortex discretisation points.
    The velocity induced on the vortex point $\svec_0$ includes the \emph{local}
    contribution \cref{eq:BS_local} of its adjacent segments ($\Ccal_0$, in orange), and
    the \emph{non-local} contribution of the rest ($\overline{\Ccal}_0$, in grey).
    Periodic images and other filaments (not depicted) also contribute to the
    non-local velocity.
    Near the evaluation point $\svec_0$, the lengths $\ell^-$ and $\ell^+$ of
    the adjacent segments are represented, as well as the local unit tangent
    $\svec_0'$ and curvature vector $\svec_0''$.
    Numerically, non-local contributions are computed by approximating line
    integrals by quadrature sums over each non-local segment.
    As an example, the open red circles between points $\svec_j$ and
    $\svec_{j + 1}$ are the quadrature points associated to that segment (here,
    using $Q = 3$ point Gauss--Legendre quadratures on each segment).
    These are the points where the integrand is actually evaluated when
    computing non-local contributions.
  }\label{fig:filament_discretisation}
\end{figure}

In practice, the VFM is used to describe length scales which are several
orders of magnitude larger than $\VortexRadius$.
In numerical simulations, vortex filaments are discretised with a typical line
resolution $\ell \gg \VortexRadius$.
Therefore, directly computing the Biot--Savart integral with a cut-off of order
$\VortexRadius$ as described above can lead to large numerical error,
especially when quadrature rules are used to estimate the integrals.
For this reason, the usual approach is to split the evaluation of the
Biot--Savart integral on a discretised filament onto
\emph{local} and \emph{non-local} contributions~\cite{Schwarz1985}, as represented in
\cref{fig:filament_discretisation} for a single filament.
The local contribution is estimated by analytically integrating the leading-order term of
the Taylor expansion \cref{eq:bs_integrand_taylor} on the local portion $\Ccal_0$ of the
discretised curve and excluding the cut-off region of length
$2\CoreSizeCutoffCoef_v \VortexRadius$.
This leads to
\begin{equation}
  \vvec\local(\svec_0) =
  \frac{\kappa}{4\pi} \svec_0' \times \svec_0''
  \left[
    \ln \left( \frac{2 \sqrt{\ell^- \ell^+}}{\VortexRadius} \right) - \Delta
  \right],
  \label{eq:BS_local}
\end{equation}
where $\ell^-$ and $\ell^+$ are the lengths of the adjacent segments composing $\Ccal_0$.
The local velocity is thus oriented along the local binormal direction and is
proportional to the local curvature $|\svec_0''|$.
As illustrated in \cref{fig:filament_discretisation}, when filaments are
discretised by connecting vortex points and $\svec_0$ is one of these points,
the adjacent segments are usually defined as those linking $\svec_0$ to its neighbours.
Finally, accounting for the local term, the total induced velocity of a point
$\svec_0$ on a discretised vortex can be written as
\begin{equation}
  \vvec(\svec_0) =
  \vvec\local(\svec_0) +
  \frac{\kappa}{4\pi} \sum_{\nvec \in \mathbb{Z}^3}
  \int_{\Ccal}' \frac{(\svec - \svec_0 + \nvec L) \times \dd \svec}{|\svec - \svec_0 + \nvec L|^3},
  \label{eq:BS_on_vortex}
\end{equation}
where the prime over the integral indicates that integration is omitted on the
local portion $\Ccal_0$ of the filament containing $\svec_0$.

\subsection{Energy estimation in periodic systems}

The VFM described above includes no dissipation mechanisms,
and energy is expected to be a conserved quantity.
It is crucial to be able to accurately estimate the total energy
of the system, as this may allow (i) to verify the accuracy of energy conservation at the
numerical level or (ii) to estimate energy decay rates when the model is extended with dissipative
mechanisms.

In a periodic domain, the kinetic energy per unit mass is defined as
\begin{equation}
  E = \frac{1}{2V} \int_{\Omega} \left| \vvec(\xvec) \right|^2 \dd^3\xvec
  \label{eq:def_kinetic_energy}
\end{equation}
where $\Omega = {[0, L]}^3$ is the main periodic cell and $V = L^3$ is its volume.
In principle, this expression requires knowing the velocity field
$\vvec(\xvec)$ at every point $\xvec$ in space.
In the VFM this is not only impractical -- as it would require many evaluations
of the Biot--Savart law \cref{eq:BS} -- but is also delicate since the velocity
field presents strong gradients near vortex filaments.
For these reasons, in non-periodic VFM simulations~\cite{Samuels2001,
Baggaley2011, Hanninen2013} the above expression is commonly replaced by\footnote{%
  Here we express the energy is per unit \emph{density} and not \emph{mass}, since the
  domain volume $V$ (and thus the fluid mass) is infinite.
}
\begin{equation}
  E_{\text{non-periodic}} = \kappa \oint \vvec(\svec) \cdot (\svec \times \dd\svec),
  \label{eq:kinetic_energy_nonperiodic}
\end{equation}
which only requires knowing the vortex geometry and the velocity of the vortex filaments.
It is based on a similar expression initially derived as a volume
integral~\cite[p. 218]{Lamb1945} (see also~\cite[p. 68]{Saffman1993}), which is
expected be valid for spatially smooth vorticity fields.
It assumes that the velocity and vorticity fields decay to zero at infinity,
which is not the case in periodic domains.
More importantly, as we shall see below, it does not correctly account for the
vortex core size $a$ and thus it is not properly conserved.

An alternative expression for the kinetic energy per unit mass can be obtained from
\cref{eq:def_kinetic_energy} using integration by parts~\cite[p. 68]{Saffman1993},
\begin{equation}
  E
  = \frac{1}{2V} \int_{\Omega} \psivec(\xvec) \cdot \vortvec(\xvec)
  \, \dd^3\xvec
  = \frac{\kappa}{2V} \oint_{\Ccal} \psivec(\svec) \cdot \dd\svec,
  \label{eq:kinetic_energy_streamfunction}
\end{equation}
where the last equality is obtained using \cref{eq:vfm_vorticity}.
Note that the boundary terms of the integration by parts vanish in periodic
domains.
In \cref{eq:kinetic_energy_streamfunction}, we have introduced the
streamfunction vector $\psivec(\xvec)$ (or \emph{vector potential}~\cite{Feynman2011, Saffman1993}), which is
related to the velocity by $\curl \psivec = \vvec$, and thus to the vorticity
by Poisson's equation $-\laplacian \psivec = \vortvec$.
In other words, the kinetic energy can be obtained from a line integral
requiring knowledge of streamfunction values on vortex filaments.
Interestingly, \cref{eq:kinetic_energy_streamfunction} also allows to interpret
the tangential streamfunction, $\psi_t = \psivec \cdot \svec'$, as the
linear energy density of a vortex point (up to a multiplicative constant
$\kappa/2V$), and thus as the contribution of a vortex element to the total
kinetic energy.
Finally, note that \cref{eq:kinetic_energy_streamfunction} is also valid in
non-periodic unbounded domains under the same assumptions leading to
\cref{eq:kinetic_energy_nonperiodic}.
To our knowledge, this expression has never been used before in the context of
the VFM\@.
Perhaps one of the reasons is that $\psivec(\svec)$ is a priori not
available in VFM simulations, and computing it comes at an additional cost.

\subsection{Obtaining the streamfunction vector}\label{sec:VFM:streamfunction}

As mentioned above, the streamfunction is the solution of the Poisson equation
$-\laplacian \psivec = \vortvec$.
In three dimensions, the solution can be explicitly written as a
convolution of $\vortvec$ with the Green's function
$G(\xvec, \yvec) = G(\xvec - \yvec) = 1 / (4\pi |\xvec - \yvec|)$,
\begin{equation}
  \psivec(\xvec)
  = (G \ast \vortvec)(\xvec)
  = \frac{\kappa}{4\pi} \sum_{\nvec \in \mathbb{Z}^3}
  \oint_{\Ccal}
  \frac{\dd \svec}{|\xvec - \svec + \nvec L|}
  \, \varphi_\psi(\xvec - \svec + \nvec L),
  \label{eq:streamfunction_integral}
\end{equation}
where once again we have used \cref{eq:vfm_vorticity} to express the vorticity
field.
In analogy with \cref{eq:BS}, here $\varphi_\psi$ accounts for the finite radius $\VortexRadius$ of the vortex core.
Noting that $\gradient G(\rvec) = -\rvec / (4\pi r^3)$, where $r = |\rvec|$, one can readily show
that taking the curl of \cref{eq:streamfunction_integral} leads to the
Biot--Savart law \cref{eq:BS} when $\xvec \notin \Ccal$ (in which case $\varphi_\psi = \varphi_v = 1$ everywhere).

Similarly to the velocity (\cref{sec:desingularisation_velocity}),
one must properly define the regularisation term $\varphi_\psi$ in
\cref{eq:streamfunction_integral} to avoid the integral from diverging when
evaluating it on a filament location $\xvec = \svec_0 \in \Ccal$.
As for the velocity, we define it as
$\varphi_\psi(\rvec) = 0$ for $|\rvec| < \CoreSizeCutoffCoef_\psi \VortexRadius$, and
$1$ otherwise.
Crucially, the cut-off coefficient $\CoreSizeCutoffCoef_\psi$ need not be
equal to the coefficient $\CoreSizeCutoffCoef_v = e^{\Delta} / 2$ used for the
velocity.
In fact, setting it to $\CoreSizeCutoffCoef_\psi = e^{\Delta - 1} / 2$ allows
to identify the resulting energy \cref{eq:kinetic_energy_streamfunction} with a
Hamiltonian for the vortex system as argued at the end of this section.
As verified numerically in \cref{sec:numerical:conservation}, this Hamiltonian
is accurately conserved in non-dissipative VFM simulations.

Using the same notation as in \cref{eq:bs_integrand_taylor}, the integrand
of \cref{eq:streamfunction_integral} behaves close to $\svec_0 = \svec(\xi_0)$ as
\begin{equation}
  \frac{\svec'(\xi)}{|\svec(\xi) - \svec_0|} \, \dd\xi
  = \frac{\svec'_0}{|\xi - \xi_0|} \, \dd\xi + O(1).
  \label{eq:streamfunction_integrand_taylor}
\end{equation}
Then, in analogy with the local velocity \cref{eq:BS_local},
integrating over
$\xi \in
[\xi_0 - \ell^-, \xi_0 - \CoreSizeCutoffCoef_\psi \VortexRadius] \cup
[\xi_0 + \CoreSizeCutoffCoef_\psi \VortexRadius, \xi_0 + \ell^+]$
($\Ccal_0$ segments in \cref{fig:filament_discretisation}) leads to the local contribution
\begin{equation}
  \psivec\local(\svec_0) =
  \frac{\kappa \svec_0'}{2\pi} \left[
    \ln \left( \frac{2 \sqrt{\ell^- \ell^+}}{\VortexRadius} \right) - (\Delta - 1)
  \right].
  \label{eq:streamfunction_local}
\end{equation}
Interestingly, this term is tangent to the filament, and therefore it
fully contributes to the kinetic energy \cref{eq:kinetic_energy_streamfunction}.
In the context of vortex dynamics, this local contribution is commonly referred to as
\emph{vortex tension}~\cite{Moore1972, Barenghi2023}.
Finally, in analogy with \cref{eq:BS_on_vortex}, the streamfunction vector on a
discretised filament can be written as
\begin{equation}
  \psivec(\svec_0) =
  \psivec\local(\svec_0) +
  \frac{\kappa}{4\pi} \sum_{\nvec \in \mathbb{Z}^3}
  \int_{\Ccal}' \frac{\dd \svec}{|\svec - \svec_0 + \nvec L|},
  \label{eq:streamfunction_on_vortex}
\end{equation}
where once again the prime over the integral denotes the omission of local
vortex elements around $\svec_0$ ($\Ccal_0$ segments in \cref{fig:filament_discretisation}).

To justify the above choice of $\CoreSizeCutoffCoef_\psi$, we consider a
circular vortex ring of radius $R \gg \VortexRadius$.
From \cref{eq:BS} (see also \cite{Schwarz1985}), its self-induced
translational velocity is
\begin{equation}
  V = \frac{\kappa}{4\pi R} \left[ \ln \left( \frac{8R}{\VortexRadius} \right)  - \Delta \right].
  \label{eq:vortex_ring_velocity}
\end{equation}
The total energy per unit mass $E$, satisfying Hamilton's equation $V = \partial E / \partial
p$ where $p = (\kappa / V) \pi R^2$ is the vortex ring impulse, is
then given by~\cite{Roberts1970, Sullivan2008}
\begin{equation}
  E = \frac{\kappa^2 R}{2V} \left[
    \ln \left( \frac{8R}{\VortexRadius} \right) - (\Delta + 1)
  \right].
  \label{eq:vortex_ring_energy}
\end{equation}
One can analytically show that $\CoreSizeCutoffCoef_\psi = e^{\Delta - 1} / 2$ is
the only possible choice allowing to obtain
\cref{eq:vortex_ring_energy} from expression \cref{eq:kinetic_energy_streamfunction}.
Later in \cref{sec:numerical:conservation}, we verify that
this choice leads to proper energy conservation up to numerical accuracy.
This appears to be the case not only for circular vortex rings, but also for
complex configurations containing multiple closed vortices of arbitrary shape,
and even unclosed vortices extending to infinity (as in \cref{fig:visu_cases}, centre),
in which case energy computations have been deemed impractical in the
past~\cite{Baggaley2011}.
Finally, note that the alternative energy expression \cref{eq:kinetic_energy_nonperiodic}
commonly used in open non-periodic systems is not consistent with Hamilton's
equation for a circular ring, as it corresponds to replacing the $\Delta + 1$
term in the ring energy \cref{eq:vortex_ring_energy} with $\Delta$.
This suggests that \cref{eq:kinetic_energy_nonperiodic} does not properly account for the
structure of the vortex core.

\subsection{Discretisation of spatial curves and line integrals}%
\label{sec:discretisation_quadratures}

We finish this section by discussing the approach we adopt to represent vortex
filaments and estimate line integrals in numerical simulations.
In the context of vortex filament simulations for describing quantum fluids,
filaments have been almost invariably described by a
set of points in space connected by straight segments~\cite{Schwarz1985,
Samuels1992, Baggaley2012f, Hanninen2013, Yui2021}.
Integration of the Biot--Savart law over lines is then straightforward and can
even be done analytically over each straight segment.
Curve derivatives (local tangents, curvatures) are estimated on the
discrete points using e.g.\ finite difference approximations~\cite{Baggaley2012f}.
While this approach can be implemented with relative ease, it is a low-order
discretisation, often requiring very small distances between discretisation points
to achieve accurate results.
It also presents an inconsistency between the $C^0$ continuity of the
lines and the $C^2$ continuity required to estimate curvatures.

Here we take a different route and consider the filaments as smooth curves
passing through a set of points, as illustrated in \cref{fig:filament_discretisation}.
As introduced in \cref{sec:filaments_as_oriented_curves},
each curve is parametrised as $\svec(\tparam)$ for some scalar parameter
$\tparam \in [0, \Tparam]$.
The numerical degrees of freedom are the positions of discrete vortex points or
\emph{nodes}.
Some interpolation method is then applied to evaluate vortex positions
in-between nodes.
The interpolation also gives direct access to curve derivatives along vortex lines,
allowing to estimate local tangent and curvature vectors required for
Biot--Savart computations.
In the numerical experiments of \cref{sec:numerical_experiments} we use quintic
spline interpolations which have global continuity $C^4$ (see \cref{sm:sec:filament_discretisation} for details).

In the context of the Biot--Savart problem, a second difficulty is the need to
estimate integrals on vortex lines.
Here we estimate line integrals using Gauss--Legendre
quadratures on each segment connecting two neighbouring nodes.
Concretely, if one considers a closed curve $\Ccal$ parametrised as
$\svec(\tparam)$ with nodes
$\left\{ \svec_j = \svec(\tparam_j); j = 1, \ldots, N \right\}$,
then an integral over such curve is approximated as
\begin{equation}
  \int_{0}^{\Tparam} F[\svec(\tparam)] \, \dd\tparam
  \approx
  \sum_{j = 1}^N \sum_{i = 1}^Q
  w_i F \big[{\svec_j^{(i)}}\big] \Delta \tparam_j,
  \quad
  \svec_j^{(i)} \equiv \svec(\tparam_j + h_i \Delta \tparam_j),
  \label{eq:line_integral_quadrature}
\end{equation}
where $Q$ is the number of quadrature points per segment (typically one chooses
$Q \le 4$ in simulations).
Here, $F[\svec]$ is a functional which may depend on curve locations
$\svec(\tparam)$ as well as its derivatives.
For convenience, the quadrature rule above is defined in the domain $[0, 1]$,
such that $h_i \in [0, 1]$ are the
quadrature locations, and $w_i$ the associated
weights satisfying $\sum_{i = 1}^Q w_i = 1$.
Moreover, $\Delta \tparam_j = \tparam_{j + 1} - \tparam_{j}$ is the parameter
increment associated to a single segment, with the convention that
$\tparam_{N + 1} = \tparam_1 + \Tparam$ for a closed or an infinite periodic curve.
As an example, the open red circles in \cref{fig:filament_discretisation} are
the quadrature points associated to a single filament segment (using $Q = 3$).

\section{Fast Ewald summation for the Biot--Savart problem}\label{sec:ewald_bs}

In this section we adapt fast Ewald summation methods to the evaluation of
velocity and streamfunction fields induced by a set of vortex filaments.
This requires some adjustments since these methods generally deal with a scalar-valued
source term (e.g.\ electrostatic charge) supported on discrete points, while
here the source term (vorticity) is vector-valued and defined on spatial curves.
In the context of the VFM, the basic idea of these methods is to split the line integrals in
\cref{eq:BS_on_vortex,eq:streamfunction_on_vortex} onto
short- and long-range components.
Concretely, including the local terms appearing in those expressions, the streamfunction and
the velocity at a vortex location $\svec_0$ are decomposed as
\begin{align}
  \psivec(\svec_0)
  &= \psivec\local(\svec_0) + \psivec\shortrange(\svec_0)
  + \psivec\longrange(\svec_0)
  - \psivec\longrange\self(\svec_0),
  \label{eq:ewald_streamfunction_splitting}
  \\
  \vvec(\svec_0)
  &= \vvec\local(\svec_0) + \vvec\shortrange(\svec_0)
  + \vvec\longrange(\svec_0)
  - \vvec\longrange\self(\svec_0),
  \label{eq:ewald_velocity_splitting}
\end{align}
where the $\shortrangeText$ and $\longrangeText$ superscripts denote near- and far-field components,
and $\psivec\longrange\self$ and $\vvec\longrange\self$ are corrections to
the latter which are discussed further below.

\subsection{Ewald splitting}

Ewald summation methods split the singular Green's function
$G(\rvec)$ associated to Poisson's equation into (i) a fast decaying part
$G\shortrange(\rvec)$ accounting for short-range interactions and (ii) a slowly
decaying part $G\longrange(\rvec)$, which is well-behaved at $\rvec = \zerovec$ and
describes long-range interactions.
In three dimensions, this is usually achieved via the identity $\erf(x) + \erfc(x) = 1$,
\begin{equation}
  G(\rvec) = \frac{1}{4\pi r}
  = \frac{\erfc(\alpha r)}{4\pi r} + \frac{\erf(\alpha r)}{4\pi r}
  = G\shortrange(\rvec) + G\longrange(\rvec),
  \label{eq:ewald_green_functions}
\end{equation}
where $\erf(x) = (2 / \sqrt{\pi}) \int_0^x e^{-u^2} \,\dd u$
and $\erfc(x) = (2 / \sqrt{\pi}) \int_x^{\infty} e^{-u^2} \,\dd u$ are
respectively the error function and the complementary error function.
Due to linearity, the solution $\psivec = G \ast \vortvec$ to the Poisson equation $-\laplacian \psivec =
\vortvec$ is accordingly split into
$\psivec = \psivec\shortrange + \psivec\longrange$.
Similarly, the velocity field can be decomposed as
$\vvec = \curl\psivec = \vvec\shortrange + \vvec\longrange$.
Importantly, $\alpha$ in \cref{eq:ewald_green_functions} is the Ewald splitting
parameter, which defines an inverse length scale setting the transition
between short- and long-range interactions.
This is a purely numerical parameter, as the physical fields $\psivec$
and $\vvec$ obtained by adding both contributions are in theory independent of $\alpha$.
To facilitate the interpretation of the splitting
\cref{eq:ewald_green_functions}, it is helpful to consider the \enquote{far-field}
vorticity field defined by
$\vortvec\longrange = -\laplacian \psivec\longrange$, which can be written as
the convolution of $\vortvec$ with $-\laplacian G\longrange$.
In fact this is a Gaussian-filtering operation, since $-\laplacian G\longrange(\rvec) = {(\alpha
/ \sqrt{\pi})}^3 e^{-{(\alpha r)}^2}$.
In other words, $\psivec\longrange$ is the streamfunction associated to a
coarse-grained (smoothed) version of the original (singular) vorticity $\vortvec$.
Similarly, $\vvec\longrange = \curl\psivec\longrange$ is the corresponding
coarse-grained velocity field.

Taking the gradient of \cref{eq:ewald_green_functions},
one arrives to the corresponding splitting for
the Biot--Savart kernel $\gradient G$,
\begin{equation}
  \gradient G(\rvec)
  = -\frac{\rvec}{4\pi r^3}
  = -\left[ g\shortrange(\rvec) + g\longrange(\rvec) \right] \frac{\rvec}{4\pi r^3}
  = \gradient G\shortrange(\rvec) + \gradient G\longrange(\rvec)
  \label{eq:ewald_green_function_derivatives}
\end{equation}
with the weight functions
\begin{equation}
  g\shortrange(\rvec) =
  \erfc(\alpha r) + \frac{2 \alpha r}{\sqrt{\pi}} \, e^{-{(\alpha r)}^2},
  \quad
  g\longrange(\rvec) =
  \erf(\alpha r) - \frac{2 \alpha r}{\sqrt{\pi}} \, e^{-{(\alpha r)}^2}.
  \label{eq:ewald_derivative_weight_functions}
\end{equation}
It can be verified that the long-range kernels are non-singular, behaving as
$G\lr(\rvec) = \alpha / ({2\pi^{3/2}}) + O(r^2)$ and
$\gradient G\lr(\rvec) = -\alpha^3 \rvec / ({3 \pi^{3/2}}) + O(r^3)$ near $r = 0$.
On the other hand, the short-range kernels asymptotically decay as
$G\sr(\rvec) \sim e^{-{(\alpha r)}^2} / \left[ 4\pi r (\alpha r)^{3/2} \right]$
and
$\gradient G\sr(\rvec) \sim \alpha \rvec \, e^{-{(\alpha r)}^2} / \left( 2 \pi^{3/2} r^2 \right)$
for large $r$.

\subsection{Estimation of short-range interactions}\label{sec:ewald_short_range}

The near-field streamfunction $\psivec\sr$ is obtained by replacing $G(\rvec)$ in
\cref{eq:streamfunction_integral} with $G\shortrange(\rvec)$ defined in
\cref{eq:ewald_green_functions}, resulting in
\begin{equation}
  \psivec\shortrange(\svec_0)
  = \frac{\kappa}{4\pi} \sum_{\nvec \in \mathbb{Z}^3}
  \int_{\Ccal}' f\shortrange(\svec - \svec_0 + \nvec L) \, \frac{\dd \svec}{|\svec - \svec_0 + \nvec L|},
  \label{eq:streamfunction_shortrange}
\end{equation}
where $f\shortrange(\rvec) = \erfc(\alpha r)$.
We recall that the prime over the integral symbol denotes the omission of local vortex
elements around $\svec_0$ ($\Ccal_0$ segments in \cref{fig:filament_discretisation}).
Similarly, following the decomposition
\cref{eq:ewald_green_function_derivatives}, the near-field velocity field is
explicitly given by the modified Biot--Savart integral
\begin{equation}
  \vvec\shortrange(\svec_0)
  = \frac{\kappa}{4\pi}
  \sum_{\nvec \in \mathbb{Z}^3}
  \int_{\Ccal}' g\shortrange(\svec - \svec_0 + \nvec L) \, \frac{(\svec - \svec_0 + \nvec L) \times \dd \svec}{|\svec - \svec_0 + \nvec L|^3}.
  \label{eq:velocity_shortrange}
\end{equation}
The original integrals in \cref{eq:BS_on_vortex,eq:streamfunction_on_vortex}
are recovered by setting $\alpha = 0$ in the above expressions.
In practice, we approximate the above line integrals using quadrature sums over discrete
vortex line locations $\svec_j^{(i)}$ according to
\cref{eq:line_integral_quadrature}.
Moreover,
since the short-range kernels $G\shortrange$ and $\gradient G\sr$ decay exponentially with $(\alpha r)^2$,
one can safely define a cut-off distance $\rcut$ beyond which short-range
interactions can be neglected.
The truncation errors associated to the choice of $\rcut$ are discussed in \cref{sec:truncation_errors}.

\subsection{Estimation of long-range interactions}\label{sec:ewald_longrange}

Since the long-range Green's function $G\lr$ is non-singular,
one can use a truncated Fourier series
representation to indirectly solve
$-\laplacian \psivec\longrange = \vortvec\longrange$.
We start by writing the periodic vorticity field as
\begin{equation}
  \vortvec(\xvec)
  \approx \sum_{\kvec \in \mathbb{K}^3} \hat{\vortvec}_{\kvec} \, e^{\im \kvec \cdot \xvec}
  \quad\text{where}\quad
  \hat\vortvec_{\kvec} = \frac{1}{V} \int_{\Omega} \vortvec(\xvec) \, e^{-\im \kvec \cdot \xvec} \, \dd^3\xvec.
  \label{eq:vorticity_fourier_series}
\end{equation}
Here $\mathbb{K} = \left\{ \frac{2\pi m}{L};
m = -\left\lfloor \frac{M}{2} \right\rfloor, \ldots, \left\lfloor \frac{M - 1}{2} \right\rfloor \right\}$
is the set of $M$ resolved Fourier wavenumbers in each Cartesian direction and
$\lfloor \cdot \rfloor$ denotes the floor operation (for simplicity, we take
$M$ to be the same in all directions).
We recall that $\Omega = {[0, L]}^3$ represents the main periodic cell and $V = L^3$ is its volume.
Since the vorticity \cref{eq:vfm_vorticity} is singular and supported on
spatial curves, its Fourier coefficients are then
\begin{equation}
  \hat\vortvec_{\kvec}
  = \frac{\kappa}{V} \oint_{\Ccal} e^{-i \kvec \cdot \svec} \, \dd\svec
  \approx \frac{\kappa}{V} 
  \sum_{l = 1}^{\Nfil}
  \sum_{j = 1}^{N_l} \sum_{i = 1}^Q
  w_i \, \Delta \tparam_{jl} \, \svec'_{ijl} \, e^{-\im \kvec \cdot \svec_{ijl}},
  \label{eq:vorticity_fourier_coefs}
\end{equation}
where the last expression is obtained from the quadrature approximation
\cref{eq:line_integral_quadrature}.
The outermost sum is over the $\Nfil$ vortex filaments of the system, each
being discretised by a possibly different number of nodes $N_l$.
Moreover, $\svec_{ijl}$ is a shorthand for $\svec_l(\tparam_j + h_i \Delta
\tparam_j)$, where $\svec_l(\tparam)$ is the parametrisation of the $l$-th filament,
while $\svec_{ijl}'$ is the derivative with respect to $\tparam$ at that location
(aligned with the local tangent vector).
The triple sum in \cref{eq:vorticity_fourier_coefs}
can be interpreted as a sum of vector charges
$\qvec_{ijl} = w_i \, \Delta\tparam_{jl} \, \svec'_{ijl}$ on locations $\svec_{ijl}$.
Note that using the same quadrature nodes for short- and long-range
computations means that interpolated values of $\svec$ and $\svec'$ can be
shared among both components, reducing the computational cost associated to interpolations.

Now, since $\vortvec\longrange$ is the convolution of $\vortvec$ with a Gaussian kernel, its
Fourier coefficients are
$\hat{\vortvec}_{\kvec}\longrange = \hat{\vortvec}_{\kvec} \, e^{-k^2 / 4 \alpha^2}$ where $k = |\kvec|$.
Unlike $|\hat{\vortvec}_{\kvec}|$, the amplitudes
$|\hat{\vortvec}_{\kvec}\longrange|$ can be expected to decay quickly for
$k \gg \alpha$, which justifies the truncation of the Fourier series.
Moreover, solving Poisson's equation amounts to division by $k^2$ in Fourier space.
This ultimately allows us to express the far-field velocity in physical space as
\begin{equation}
  \vvec\longrange(\xvec)
  \approx \sum_{\substack{\kvec \in \mathbb{K}^3 \\ |\kvec| \ne 0}}
  \hat\vvec\longrange_{\kvec} \, e^{\im \kvec \cdot \xvec}
  \quad\text{with}\quad
  \hat\vvec\longrange_{\kvec}
  = \im \kvec \times \hat\psivec\lr_{\kvec}
  = \im \kvec \times \frac{\hat\vortvec\lr_{\kvec}}{k^2}.
  \label{eq:velocity_longrange_physical}
\end{equation}
Note that this requires
$\hat{\vortvec}_{\zerovec} = \zerovec$, i.e.\ the mean vorticity within a periodic
cell must be zero.
The far-field streamfunction field $\psivec\longrange(\xvec)$ can be written
similarly to \cref{eq:velocity_longrange_physical}
from its Fourier coefficients $\hat\psivec\longrange_{\kvec} = \hat\vortvec\longrange_{\kvec} / k^2$.

When evaluated on a vortex location $\svec_0$, the obtained far-field velocity
and streamfunction vectors include contributions of the local segments
adjacent to $\svec_0$.
However, integration over these segments must
be excluded according to \cref{eq:BS_on_vortex,eq:streamfunction_on_vortex}.
This issue is analogous to the spurious self-interaction electrostatic potential
appearing in standard Ewald methods~\cite{Arnold2013}.
In the present case, the spurious local integrals are explicitly given by
\begin{align}
  \psivec\longrange\self(\svec_0)
  &= \frac{\kappa}{4\pi}
  \int_{\Ccal_0} f\longrange(\svec - \svec_0) \, \frac{\dd \svec}{|\svec - \svec_0|},
  \label{eq:streamfunction_longrange_self}
  \\
  \vvec\longrange\self(\svec_0)
  &= \frac{\kappa}{4\pi}
  \int_{\Ccal_0} g\longrange(\svec - \svec_0) \frac{(\svec - \svec_0) \times \dd \svec}{|\svec - \svec_0|^3},
  \label{eq:velocity_longrange_self}
\end{align}
where $f\longrange(\rvec) = \erf(\alpha r)$, $g\longrange(\rvec)$ is defined
in \cref{eq:ewald_derivative_weight_functions},
and $\Ccal_0$ consists of the two segments adjacent to $\svec_0$ (see
\cref{fig:filament_discretisation}).
They can be estimated using a variant of the quadrature procedure
\cref{eq:line_integral_quadrature}, thus requiring $2Q$ evaluations of the
integrand for each vortex location $\svec_0$.
As indicated by
\cref{eq:ewald_streamfunction_splitting,eq:ewald_velocity_splitting}, these two
\enquote{self-interaction} terms must be subtracted from the long-range estimations
to avoid an unphysical dependence of the results on the Ewald parameter $\alpha$.

As in previous works~\cite{Hedman2006, Arnold2013, Pippig2013}, here we adopt
the NUFFT algorithm~\cite{Dutt1993} to
efficiently approximate the Fourier sums \cref{eq:vorticity_fourier_coefs} and
\cref{eq:velocity_longrange_physical}.
These can be respectively obtained using type-1
(non-uniform to uniform) and type-2 (uniform to non-uniform) NUFFTs~\cite{Greengard2004}.
When relying on the NUFFT, the accuracy of long-range computations is then controlled by
the internal NUFFT parameters and the truncation wavenumber
$\kmax = (2\pi / L) \lfloor (M - 1) / 2 \rfloor$.
These two sources of error are respectively discussed in \cref{app:nufft}
and \cref{sec:truncation_errors}.

\section{Truncation error estimates}\label{sec:truncation_errors}

We now provide estimates of the root-mean-square errors associated to the
short- and long-range cut-offs $\rcut$ and $\kmax$.
Similar estimates have already been provided for the electrostatic
problem~\cite{Kolafa1992, Deserno1998a}, but these do not directly
apply to the VFM since (i) the quantities of interest are not the same, and
(ii) the singular sources in the present case are spatial curves and not
points.
We also show that the effect of the parameters $\alpha$,
$\rcut$ and $\kmax$ on accuracy can be reduced to a unique non-dimensional coefficient
$\beta$, thus greatly
simplifying the parameter selection procedure.

The derivation of the error estimates is detailed in
\cref{sm:sec:truncation_errors_derivation}.
In summary, the estimates make the simplifying assumptions that (i) vortex
filaments are homogeneously distributed in the spatial domain, (ii) the
relative orientation of two vortex elements at a distance $r > \rcut$ is
completely decorrelated, and (iii) $\kmax$ is much larger than the typical
curvature $|\svec''|$ of the vortices.
Introducing the non-dimensional cut-off coefficients $\beta\sr = \alpha \rcut$
and $\beta\lr = \kmax / 2\alpha$, we show that both short- and long-range
errors decay exponentially with the square of these respective coefficients,
which justifies reducing them into a unique coefficient $\beta$.
This finally leads to the truncation error estimates
\begin{align}
  \label{eq:error_trunc_velocity}
  \abserr_v
  &\approx
  \kappa \left[
    \frac{\VortexLength}{\sqrt{\pi} \alpha V}
    +
    \frac{1}{\beta} {\left( \frac{\VortexLength}{8\pi V} \right)}^{\!\! 1/2}
  \right]
  e^{-\beta^2},
  \\
  \label{eq:error_trunc_streamfunction}
  \abserr_\psi
  &\approx
  \frac{\kappa}{2\alpha}
  \left[
    \frac{\VortexLength}{\alpha V}
    +
    \frac{1}{\beta^{1/2}}
    {\left( \frac{\VortexLength}{8\pi V} \right)}^{\!\! 1/2}
  \right]
  \frac{e^{-\beta^2}}{\beta^{3/2}},
\end{align}
where the short- and long-range contributions to the errors correspond to the
first and second terms within each square bracket.
The accuracy of the method is thus mainly driven by a unique
non-dimensional parameter $\beta$.
This sets both physical- and Fourier-space cut-offs $\rcut = \beta/\alpha$ and
$\kmax = 2\beta\alpha$ for a
given value of the inverse splitting distance $\alpha$, which is left as a free
parameter that can be adjusted to optimise performance (\cref{sec:numerical:performance}).

\section{Numerical experiments}\label{sec:numerical_experiments}

We now perform different numerical experiments to evaluate the accuracy,
conservation properties and computational complexity of the proposed FFT-based
method.
All computations are performed in double precision arithmetic.
The methods are implemented in the open-source VortexPasta.jl
package\footnote{Available at
\url{https://github.com/jipolanco/VortexPasta.jl} under the MPL-2.0 license.}~\cite{VortexPasta} written in Julia~\cite{Bezanson2017}.
The online documentation includes installation instructions, usage examples and
tutorials, a description of the model and numerical methods, hints for
running on computing clusters and graphical processing units (GPUs), and an extensive API reference.
This should allow a new user to quickly get started with running simulations
even without any prior knowledge of the Julia language.

In all numerical tests below, the domain is periodic with period $L = 2\pi$ in
all directions.
The vortex core size and the vorticity profile parameter  -- both appearing in the
locally induced velocity \cref{eq:BS_local} -- are respectively set to
$\VortexRadius = 10^{-8}$ and $\Delta = 1/4$, and the vortex circulation is set to $\kappa = 1$.
Vortex lines are represented as smooth curves using quintic splines,
which allows to estimate derivatives
and evaluate quantities in-between discretisation points.
Moreover, line integrals are estimated using $Q = 3$ quadrature points per
vortex segment (see \cref{sec:discretisation_quadratures}).

\subsection{Accuracy}\label{sec:numerical:accuracy}

\begin{figure}
  \begin{center}
    \includegraphics[width=\textwidth]{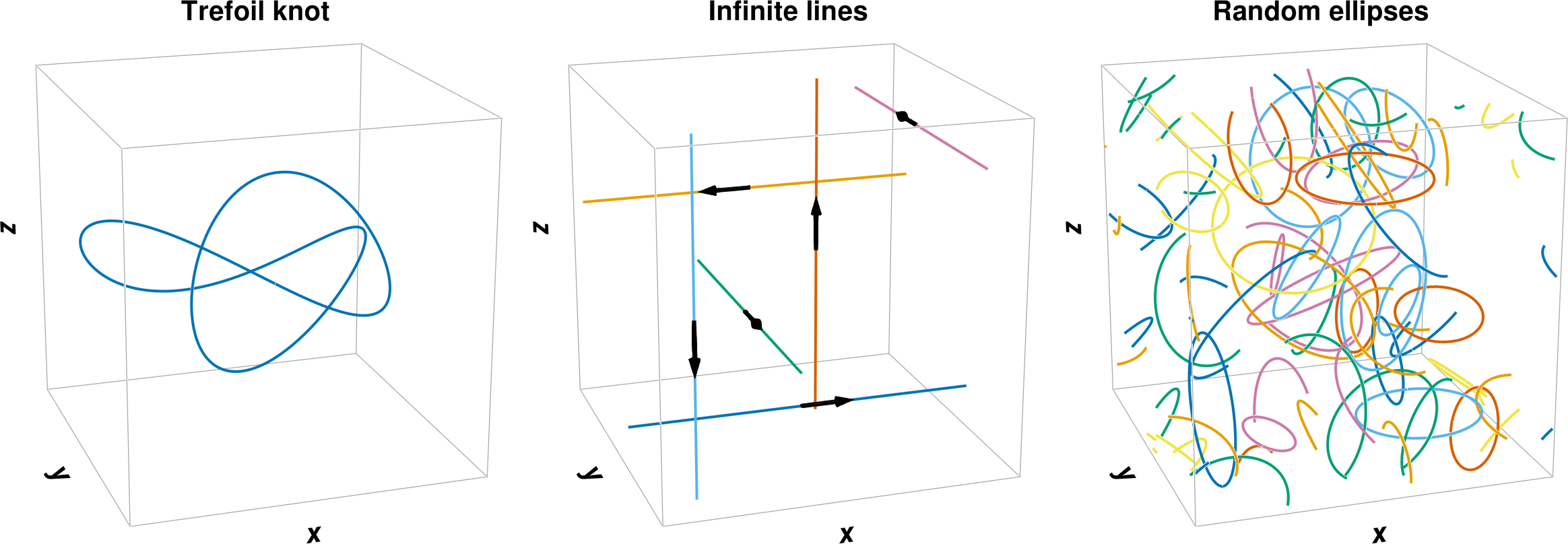}
  \end{center}
  \caption{%
    Visualisation of considered test cases within a periodic cell of size $L^3 = (2\pi)^3$.
    Left: trefoil knot with characteristic size $R = \pi/3$.
    Centre: infinitely extended straight lines.
    Arrows indicate the line orientation.
    Right: \NumberOfEllipses{} ellipses with random locations, orientations,
    sizes and aspect ratios.
    For visualisation purposes, curves crossing the domain boundaries are
    broken onto multiple parts, which are periodically wrapped to fit within
    the periodic cell.
    Colours allow to distinguish between individual vortex lines.
  }\label{fig:visu_cases}
\end{figure}

To numerically assess the accuracy of the proposed method,
we consider three different test cases of varying complexity.
The test cases are detailed below and illustrated in \cref{fig:visu_cases}.

\paragraph{1.~Trefoil knot}
The trefoil knot consists in a single knotted curve
(\cref{fig:visu_cases}, left), interacting with itself and its periodic
images.
The trefoil knot curve is parametrically defined by
\begin{equation}
  \Xvec(\theta) = R \left[
    \begin{array}{c}
      \sin(\theta) + 2 \sin(2\theta) \\
      \cos(\theta) - 2 \cos(2\theta) \\
      -\sin(3\theta)
    \end{array}
  \right]
  \quad\text{for } \theta \in [0, 2\pi],
  \label{eq:def_trefoil_knot}
\end{equation}
where $R$ determines the size of the trefoil knot.
We take $R = L/6$ so that periodicity effects are non-negligible.
The curve is discretised by evaluating \cref{eq:def_trefoil_knot} on $N = 512$
equally spaced values of $\theta$.

\paragraph{2.~Infinite lines}

This test case consists of a set of straight infinitely extended lines inducing
a three-dimensional velocity field.
Each periodic cell is crossed by three pairs of lines aligned positively and
negatively with each Cartesian direction (\cref{fig:visu_cases}, centre).
The filament orientations are such that the velocity induced on each line is non-zero
and positively aligned with the line ($\vvec \cdot \svec' > 0$).
Since the lines have zero curvature, the velocity induced by a line on itself
is zero, and therefore the velocity of a vortex point is completely due to non-local
interactions.
Therefore, this test case emphasises the accuracy of long-range computations.
We discretise each filament with $N_l = 128$ independent nodes (for $l = 1, \ldots,
6$).

\paragraph{3.~Random ellipses}

The third test case is closer to a disordered configuration relevant to
turbulent vortex flows.
It consists of \NumberOfEllipses{} ellipses which are
randomly positioned and oriented within the spatial domain (\cref{fig:visu_cases}, right).
The size and aspect ratio of each ellipse is also random.
Concretely, the minor and major radii are random
values uniformly and independently distributed in $[L/16, L/4]$.
We discretise each ellipse with $N_l = 128$ nodes (for $l = 1, \ldots, \NumberOfEllipses$).

\subsubsection*{Empirical relative errors}

To numerically quantify the truncation errors in each test case, we start by
computing a reference solution with very high accuracy.
Concretely, we set the cut-off coefficient to $\beta = 8$ and compute NUFFTs
with a nominal relative tolerance $\sim 10^{-14}$.
Here, the reference solution corresponds to a set of velocity and
streamfunction values, $\vvec_i\reference$ and $\psivec_i\reference$, on each discrete filament
node $\svec_i$.
We then compare this reference solution with the results $\vvec_i$ and
$\psivec_i$ obtained by varying $\beta$.
The splitting parameter is set to $\alpha = 24/L$
in all numerical experiments.
Accuracy estimates are obtained by evaluating the relative $\ell_2$
errors
\begin{equation}
  \relerr_{v}
  = \frac{\lVert \vvec - \vvec\reference \rVert}{\lVert \vvec\reference \rVert}
  \quad\text{ and }\quad
  \relerr_{\psi}
  = \frac{\lVert \psivec - \psivec\reference \rVert}{\lVert \psivec\reference \rVert},
  \label{eq:def_relative_errors}
\end{equation}
where the $\ell_2$ norm of a vector field $\uvec$ evaluated at $N$ discrete
locations is defined by ${\lVert \uvec \rVert}^2 = \frac{1}{N} \sum_{i = 1}^{N} |\uvec_i|^2$.
Here $|\cdot|$ denotes the vector magnitude.

\begin{figure}
  \begin{center}
    \includegraphics[]{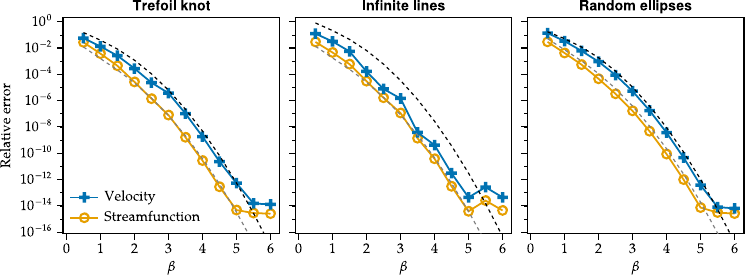}
  \end{center}
  \caption{%
    Accuracy of full Biot--Savart computations for three different test cases.
    Relative root-mean-square error of
    filament velocity $(+)$ and streamfunction values $(\times)$
    at varying non-dimensional cut-off parameter $\beta$.
    The reference solution is obtained using $\beta = 8$.
    The Ewald splitting parameter is kept constant, $\alpha = 24 / L$.
    The black dashed lines represent the total velocity error estimate
    \cref{eq:error_trunc_velocity}.
    The grey dashed lines represent the total streamfunction error estimate
    \cref{eq:error_trunc_streamfunction}.
  }\label{fig:accuracy_both_components}
\end{figure}

The obtained relative errors are plotted in \cref{fig:accuracy_both_components} (lines with symbols).
In almost all cases, the relative error decays exponentially with the squared
cut-off parameter $\beta^2$, until roundoff errors become dominant.
The decays are in remarkably good agreement with the rms error estimates
\cref{eq:error_trunc_velocity,eq:error_trunc_streamfunction} (dashed lines),
which in almost all cases provide upper bounds to the actual errors.
This confirms the relevance of the truncation estimates in a variety of configurations.

\subsection{Timestepping and conservation properties}\label{sec:numerical:conservation}

The VFM as described in \cref{sec:VFM} is expected to conserve the total
energy \cref{eq:kinetic_energy_streamfunction} and the hydrodynamic impulse
$\pvec = \frac{\kappa}{2V} \oint_{\Ccal} \svec \times \dd\svec$.
To verify this, we perform a VFM simulation in which
the position of each vortex point $\svec$ evolves in time according to
$\deriv{\svec(t)}{t} = \vvec(\svec(t), t)$.
Here the right hand side is the
Biot--Savart velocity \cref{eq:BS} evaluated at location $\svec(t)$ and at time $t$.
The test case we consider here is inspired by the classical problem of two
coaxial circular vortex rings which travel in the same direction.
Due to their mutual interaction, the rings are expected to continuously change
their sizes and therefore their translational velocities.
As a result, they pass through one another in a cyclic fashion in
\enquote{leapfrogging} motion.
This phenomenon has received much attention in classical viscous
fluids~\cite{Shariff1992} and has also been investigated in VFM simulations~\cite{Wacks2014}.

\begin{figure}
  \begin{center}
    \includegraphics[width = \textwidth]{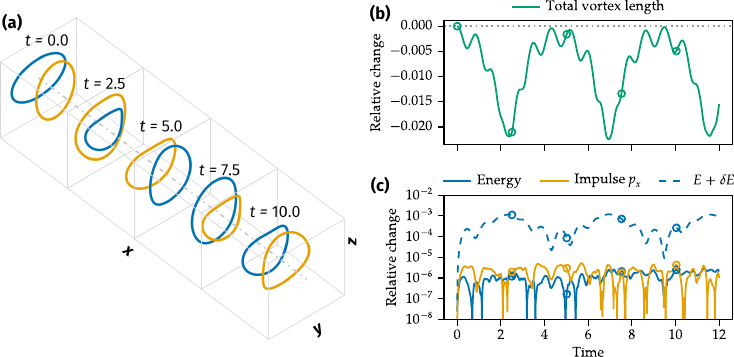}%
  \end{center}
  \caption{%
    Energy and impulse conservation of two leapfrogging elliptical rings.
    (a) Vortex geometry at different time instants.
    Vortices mainly advance in the $x$ direction, visiting 4 periodic boxes
    (light grey cubes) in the shown simulation time.
    Also shown is the symmetry axis aligned with the $x$ direction (dash-dotted line).
    (b) Relative change of the total vortex length $\VortexLength$ over time.
    (c) Absolute value of the relative change of kinetic energy $E$ and axial impulse
    $p_x$ over time.
    The dashed line corresponds to an incorrect estimation of the energy $E' = E + \delta E$
    discussed in the main text.
    Circles in (b) and (c) correspond to the snapshots in (a).
    Times are normalised by $R_a^2/\kappa$.
  }\label{fig:leapfrogging_rings}
\end{figure}

Here we consider a slightly more complex variant of the classical problem:
instead of perfectly circular vortex rings we simulate elliptical ones, so
that not only their size changes in time but also their shape.
The two elliptical vortex rings evolve in a cubic periodic domain of size $L$.
The vortices are initially located on two parallel planes -- both orthogonal to the $x$
axis -- at a distance $L/4$ from each other.
Their minor and major radii are respectively set to $R_a = L/4$ and
$R_b = R_a \sqrt{2}$.
Their size comparable to $L$ means that the effect of periodic images
cannot be fully neglected.
Besides, the ellipses are rotated relative to each other so that their major
axes are respectively aligned with the $y$ and $z$ axes.
This initial configuration is visible in \cref{fig:leapfrogging_rings}(a).

Aiming at a relative accuracy of $\sim 10^{-6}$, each vortex is discretised
with a relatively small number of points $N = 32$.
Such a coarse discretisation can lead to errors in the local terms
\cref{eq:BS_local,eq:streamfunction_local} due to the use of Taylor expansions.
For this reason, we subdivide the local filament segments
($\Ccal_0$ in \cref{fig:filament_discretisation})
so that Taylor expansions are applied on a smaller central portion (about 10\%
of the local segment length),
while 3-point Gauss--Legendre quadratures are used in the remaining
portions not in contact with the singularity.
Each discrete vortex point $\svec_i$ is evolved in time according to
$\deriv{\svec_i}{t} = \vvec(\svec_i)$.
For the temporal discretisation we adopt the standard fourth-order explicit
Runge--Kutta (RK4) scheme.
Spline coefficients are updated each time the points $\svec_i$ are advanced.
While the vortices can stretch and shrink over time, we do not perform any
remeshing of the filaments for the sake of simplicity (but this is indeed
needed in more complex problems).
To guarantee near 6-digit accuracy, we set $\beta = 3.5$
(\cref{fig:accuracy_both_components}), while the
splitting parameter is $\alpha = 7/L$ so that the short-range cut-off distance
is $\rcut = \beta/\alpha = L/2$.\footnote{%
  The condition $\rcut \le L/2$ ensures that, in short-range computations, a
  pair of vortex points sees each other at most once.
  This simplifies the implementation as one does not need to explicitly deal
  with periodic images.
  In molecular dynamics this is called the \emph{minimum image
  convention}~\cite{Arnold2005}.
}

To preserve stability, the maximum allowed timestep $\Delta t$ in VFM
simulations is roughly proportional to $\dmin^2$~\cite{Schwarz1985,
Hanninen2014} where $\dmin$ is the smallest spacing between vortex points -- in
this test case $\dmin \approx 2\pi R_a / N$.
Physically, this condition can be understood by considering a
small-amplitude sinusoidal perturbation of wavelength $\lambda$ on an otherwise
straight vortex.
The perturbation is physically expected to rotate about the vortex axis with a period
$\Tkw(\lambda) \approx \frac{2 \lambda^2}{\kappa}
{\left[ \ln\left( \frac{\lambda}{\pi
\VortexRadius} \right) + \frac{1}{2} - (\Delta + \gamma) \right]}^{-1}$,
where $\gamma \approx 0.57721$ is the Euler--Mascheroni constant~\cite{Thomson1880, Baggaley2014}.
Such periodic motion is known as a Kelvin wave.
Numerically, the timestep $\Delta t$ must therefore be small enough to capture
the Kelvin wave oscillations associated to the smallest resolved scale
$\lambda \sim \dmin$.
In practice, we find that setting $\Delta t = C \, \Tkw(\dmin)$ with $C = 1$ is
enough for the RK4 scheme to remain stable.
We finish by mentioning that the above restriction is actually imposed by the local
self-interaction term \cref{eq:BS_local}~\cite{Buttke1988, DeLaHoz2009}.
This suggests the use of splitting methods~\cite{McLachlan2002, Blanes2024} --
perhaps in combination with the Hasimoto transform~\cite{Hasimoto1972,
Banica2024} to deal with the local term -- to relax such strong restriction in
future works.

Different temporal snapshots of the simulation are visualised in
\cref{fig:leapfrogging_rings}(a).
The two vortices travel in the $x$ direction while passing through one another
as their size and shape change.
In particular, at time $t = 2.5$ the first vortex is going through the second one,
while the opposite has just occurred at time $t = 7.5$.
This cyclic motion is also visible in \cref{fig:leapfrogging_rings}(b), which
shows the temporal variation of the total vortex length $\VortexLength$.
As expected, this quantity is not conserved, displaying relatively slow
variations of up to 2\% which are clearly correlated with the physical
leapfrogging motion.
These slow variations coexist with higher-frequency fluctuations which are
explained by the non-circular shape of each vortex.
On the other hand, as seen in \cref{fig:leapfrogging_rings}(c), the total
kinetic energy \cref{eq:kinetic_energy_streamfunction} and the total
hydrodynamic impulse are approximately conserved.
Indeed, they display temporal variations uncorrelated with the leapfrogging
motion which are well within the $10^{-6}$ relative accuracy of the
calculations.

We stress that, as discussed in \cref{sec:VFM:streamfunction}, observing a
high-accuracy energy conservation rests on the proper choice of near-singularity cut-off
$\CoreSizeCutoffCoef_\psi = e^{\Delta - 1}/2$ for the
streamfunction integral \cref{eq:streamfunction_on_vortex}.
Remarkably, while this choice was justified in \cref{sec:VFM:streamfunction} for
circular vortex rings, it appears to stay valid for more complex vortex geometries.
If instead one used the same cut-off $e^{\Delta}/2$ as for the velocity
integral (exchanging $\Delta - 1$ with $\Delta$ in \cref{eq:streamfunction_local}),
this would result according to \cref{eq:kinetic_energy_streamfunction} in an
extra (negative) energy contribution $\delta E = -\frac{\kappa^2}{4\pi V} \VortexLength$.
The resulting energy $E' = E + \delta E$ would not be conserved since the
vortex length $\VortexLength$ is allowed to fluctuate in time.
This is verified in our simulations, where $E'$ (dashed line in
\cref{fig:leapfrogging_rings}(c)) displays fluctuations
which are
correlated with the fluctuations of $\VortexLength$ and are about 3 orders of
magnitude larger than those of $E$.

\subsection{Performance}\label{sec:numerical:performance}

We finally investigate the computational complexity of the proposed method
in terms of the total number of vortex points $N$.
We start by discussing how to choose the splitting parameter $\alpha$
to achieve optimal complexity for a fixed value of the accuracy parameter $\beta$.
Intuitively, $\alpha$ controls the relative amount of work done by short- and
long-range computations.
From the point of view of short-range computations, one may attempt to choose
$\alpha$ so that it reduces their expected complexity from $O(N^2)$ to $O(N)$.
Assuming a spatially homogeneous vortex point distribution, the runtime associated to
the short-range part can be estimated as $T\sr \propto Q N^2 (\rcut/L)^3$.
In terms of $\alpha$ and $\beta$, this becomes $T\sr \propto Q N^2 \beta^3 / (\alpha L)^3$.
Therefore, choosing
\begin{equation}
  \alpha(N) = C_\beta N^{1/3} / L
  \label{eq:alpha_optimal_performance}
\end{equation}
can be expected to lead to linear complexity, namely $T\sr \propto QN (\beta / C_\beta)^3$.
While $C_\beta$ should be regarded as a constant when $N$ is varied, its
optimal value may vary with the accuracy parameter $\beta$, hence the notation.
The next question is how does the choice \cref{eq:alpha_optimal_performance}
impact the complexity of long-range computations.
These are dominated by the cost of the NUFFTs, which basically consist of $\propto QN$
spreading and interpolation operations and $O(M^3 \log M)$ 3D FFTs (see
\cref{app:nufft}).
The former are already linear in $N$ and do not depend on $\alpha$.
As for the latter, the size of each 1D FFT is $M \approx \kmax L / \pi = 2\beta \alpha L / \pi$, so
that choosing $\alpha$ according to \cref{eq:alpha_optimal_performance} leads
to $M \propto \beta C_\beta N^{1/3}$ and therefore $O(N \log N)$ complexity.
In conclusion, choosing $\alpha$ as in \cref{eq:alpha_optimal_performance} can
be expected to lead, at worst, to $O(N \log N)$ complexity of the full method.

To numerically assess the performance of the method, we adapt the random
ellipses test case introduced in \cref{sec:numerical:accuracy}, illustrated in
\cref{fig:visu_cases} (right).
Each ellipse is now discretised with $N_l = 32$ vortex points, and the
Biot--Savart velocity is evaluated on all vortex points for systems with up
to 8192 vortices ($N = 2^{18}$ points).
The benchmarks are executed on a single core of
an Intel Core i7-12700H laptop processor using Julia 1.10.3.
To speed up short-range computations, a modified cell list
algorithm~\cite{Mattson1999} is used to find neighbouring points, in which the
spatial domain is divided into cells of size $\rcut / 3$ in each direction.
We consider 7 different accuracy levels, determined by $\beta$ and the NUFFT
parameters as detailed in \cref{tab:performance}.
In each case, we choose $C_\beta$ so that the time spent in short- and
long-range computations is approximately equal.
Empirically, we find that $C_\beta \approx 2.6 \beta^{-1/3}$ roughly fulfills this criterion, but
the proper choice this prefactor is likely dependent on the actual implementation
and on the machine where the tests are run.

\begin{table}
  \caption{%
    Parameters used in performance tests and summary of benchmark results.
    $\tol$, nominal relative tolerance;
    $\beta$, cut-off parameter;
    $C_\beta$, constant prefactor in \cref{eq:alpha_optimal_performance};
    $\sigma$ and $w$, NUFFT parameters (see \cref{app:nufft}).
    The last two columns summarise the obtained runtimes:
    (1) prefactor $C\fit$ of the least-squares fit
    $T\fit(N) = C\fit N \log_{10} \! N$ and
    (2) relative fitting error $\varepsilon\fit = \lVert T\fit - T \rVert / \lVert T \rVert$.
  }%
  \label{tab:performance}
  \begin{center}
    \begin{tabular}[c]{ccccc|cc}
      \toprule
      $\tol$ & 
      $\beta$ &
      $C_\beta$ &
      $\sigma$ &
      $w$ &
      $C\fit$ (\si{\micro\second}) &
      $\varepsilon\fit$ (\%)
      \\
      \midrule
      \num{e-03} & \num{2.0} & \num{2.06} & \num{1.5} & \num{4} & \num{1.41} & \num{6.07} \\
      \num{e-04} & \num{2.5} & \num{1.92} & \num{1.5} & \num{6} & \num{2.07} & \num{2.73} \\
      \num{e-06} & \num{3.5} & \num{1.71} & \num{1.5} & \num{8} & \num{4.52} & \num{2.25} \\
      \num{e-08} & \num{4.0} & \num{1.64} & \num{1.5} & \num{10} & \num{6.56} & \num{4.01} \\
      \num{e-10} & \num{4.5} & \num{1.57} & \num{1.5} & \num{12} & \num{8.95} & \num{1.81} \\
      \num{e-12} & \num{5.0} & \num{1.52} & \num{1.5} & \num{14} & \num{12.17} & \num{0.70} \\
      \num{e-14} & \num{5.5} & \num{1.47} & \num{1.5} & \num{16} & \num{15.63} & \num{0.99} \\
      \bottomrule
    \end{tabular}
  \end{center}
\end{table}

\begin{figure}
  \begin{center}
    \includegraphics{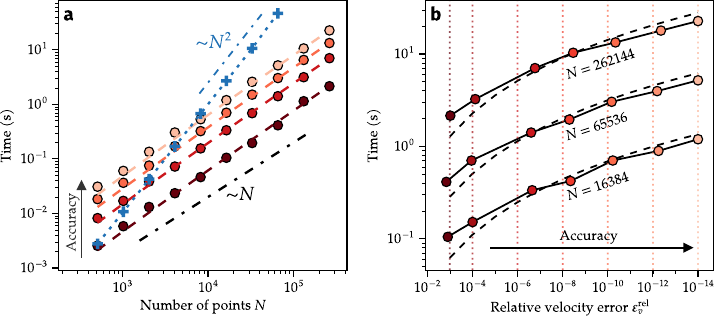}
  \end{center}
  \caption{%
    Performance of Biot--Savart velocity computations.
    (a) Runtime as a function of the number of vortex points $N$.
    Circles: proposed method for relative tolerances $\num{e-3}$, $\num{e-6}$,
    $\num{e-10}$ and $\num{e-14}$ (from darker to lighter).
    Dashed lines represent least-squares fits $T\fit(N) = C\fit N \log_{10} \! N$;
    see \cref{tab:performance} for values of $C\fit$.
    Blue crosses: naïve Biot--Savart computation in non-periodic domain.
    (b) Influence of requested accuracy on runtimes, for different values of $N$.
    Except for the rightmost circles, errors are relative to the highest
    accuracy case ($\num{e-14}$ relative tolerance).
    Dotted vertical lines represent the \emph{nominal} relative tolerances (\cref{tab:performance}).
    Dashed black lines correspond to the empirical scaling
    $T \propto p_\err^2 N \log_{10} \! N$ where $p_\err = -\log_{10} \tol$
    is the estimated number of precision digits.
  }\label{fig:performance}
\end{figure}

In \cref{fig:performance} we present the results of the numerical experiments.
Each timing is obtained as the average over 10 runs.
Panel (a) shows the runtime associated to evaluating the Biot--Savart velocity
on all vortex points for a wide range of problem sizes $N$.
The proposed method displays near linear $O(N \log N)$ complexity for all different accuracy levels
(least-square fits are detailed in \cref{tab:performance}).
For comparison, we also show the result of a naïve computation of the
Biot--Savart velocity (blue crosses), which displays the expected $O(N^2)$ scaling.
Note that the naïve computation does not account for periodic boundary
conditions, and thus the results are not directly comparable to those obtained by the
proposed method.

From \cref{fig:performance}(a), it is clear that increasing the accuracy level
can have an important impact on runtimes.
This is quantified in more detail in panel (b) for a few different
problem sizes $N$ and for all different tolerances listed in
\cref{tab:performance}.
Empirically, runtimes are seen to roughly increase as $T \propto p_\err^2$
(black dashed lines), where $p_\err = -\log_{10} \tol$ is the nominal number of
accuracy digits.
The figure also confirms that the nominal tolerances $\tol$
(vertical dotted lines) are reasonable (and in fact pessimist) estimates of the
actual errors $\relerr_v$ (markers).
To be clear, except for the $\tol = \num{e-14}$ case, the plotted errors are computed via
\cref{eq:def_relative_errors} by taking the $\tol = \num{e-14}$ case as reference.

\section{Conclusions}
\label{sec:conclusions}

We have introduced an efficient technique for the numerical evaluation of the
Biot--Savart law in periodic systems when the source field (e.g.\ current or
vorticity) is supported on three-dimensional spatial curves.
The approach is adapted from fast Ewald summation methods commonly used to
accelerate particle simulations.
The present work is relevant to applications in electromagnetism and fluid dynamics,
where the source fields respectively induce magnetic or velocity fields around the curves.
As an application, we have considered the vortex filament model (VFM)
describing superfluid helium flows near absolute zero.
In this context, the Fourier-space representation of the induced fields
is a very attractive feature of the method, as it allows the efficient
evaluation of physically relevant quantities such as energy
spectra~\cite{Barenghi2014}, coarse-grained velocity fields~\cite{Baggaley2012, Laurie2023}
and velocity circulation~\cite{Muller2021, Polanco2021}.

Several tunable parameters are introduced by the method, including an inverse
\enquote{splitting} length scale $\alpha$ and two cut-off parameters $\rcut$
and $\kmax$ in physical and Fourier spaces.
We have shown that the choice of the latter two can be reduced to that of a
single non-dimensional coefficient $\beta$ setting the accuracy of the method,
while $\alpha$ can be tuned to maximise performance.
For a fixed accuracy level, choosing $\alpha \propto N^{1/3}$ leads to near-linear $O(N
\log N)$ complexity of the method, where $N$ is the number of vortex discretisation points.
To our knowledge, this parameter selection procedure contrasts with the
seemingly common approach in molecular dynamics which consists in first
specifying reasonable cut-offs $\rcut$ and $\kmax$ in dimensional
units, and then choosing a value of $\alpha$ which maximises
accuracy given these cut-offs~\cite{Deserno1998a, Ballenegger2008, Arnold2013}.

In terms of performance, our implementation is already
capable of simulating turbulent systems with a reasonably large number
of vortices using thread-based parallelism.
Moreover, our NUFFT implementation (\cref{app:nufft}) enables the use of GPUs
to accelerate the computation of long-range interactions.
The short-range part may also benefit in the near future from a GPU implementation
of the cell lists algorithm for neighbour finding.
Besides, these core algorithms are very amenable to distributed-memory
parallelism (via MPI) using standard domain decomposition approaches, which
could enable computations on multiple CPU or GPU nodes.
Such strategies will be investigated in the future.

A second challenge concerns the timestep stability requirement $\Delta t
\propto \dmin^2$, which becomes very restrictive as the smallest resolved scale
$\dmin$ is decreased.
Within this work, this issue is partially remediated by representing spatial
curves using highly continuous splines, which allows to use relatively large
values of $\dmin$.
Potential strategies for further improving on these limitations can include the use of
splitting~\cite{Blanes2024} or multirate~\cite{Sandu2019} timestepping methods, in which the
evolution equations would be split onto a fast term responsible for
oscillations (e.g.\ the local self-induced velocity) and a
slower term (e.g.\ all non-local contributions).

The proposed method can be readily generalised to cases where the
vortex circulation $\kappa$ varies along the filaments.
In particular, it may be used in the implementation of vortex filament methods for
classical flows~\cite{Cottet2000}.
Another possible extension concerns the simpler two-dimensional (2D) case, in which
vortex filaments are replaced by point vortices.
Periodic point vortex systems have received much attention~\cite{Weiss1991,
Newton2009, vanKan2021a, Krishnamurthy2023, Grotto2024}, in part due to their
link with statistical mechanics and with 2D turbulence.
In that case, Ewald summation can be achieved using a function splitting similar to
\cref{eq:ewald_green_functions}~\cite{Hasimoto1959, Cichocki1989}.
Finally, and more directly relevant to the motivation of this work, the proposed method
may be applied to the study of \emph{finite-temperature} liquid
helium-4, in which the superfluid flow coexists and interacts with a normal viscous fluid.
A very relevant model for describing this regime consists in coupling the VFM
for the superfluid with the incompressible Navier--Stokes equations governing
the normal fluid~\cite{Kivotides2000, Yui2018, Galantucci2020}.
When the latter are solved using standard pseudo-spectral methods~\cite{Canuto1988},
the Fourier-space representation included in the present method
can enable an accurate and efficient two-way transfer of interaction forces between the
vortices and the normal fluid.

\appendix

\section*{Acknowledgments}

The author acknowledges support from the French Agence Nationale de la
Recherche through the QuantumVIW project (Grant No.\ ANR-23-CE30-0024-04).

\section{NUFFT details and influence on Biot--Savart accuracy}%
\label{app:nufft}

Here we consider the error induced by the NUFFT tolerance on the long-range
component of Biot--Savart computations.
Briefly, a type-1 NUFFT can be summarised in three steps: (i) spreading
values from a set of non-uniform locations onto a uniform grid using some
smoothing kernel, (ii) performing a regular FFT on that grid, and (iii) undoing the
spreading operation by deconvolution in Fourier space~\cite{Greengard2004}.
The type-2 NUFFT is the adjoint operation, roughly consisting of the same steps
in opposite order.
To reduce aliasing errors, the first two steps are done in a grid of $\sigma M$
points in each direction, where $M$ is the number of desired Fourier modes and
$\sigma > 1$ is an oversampling factor.
The accuracy of the NUFFT is then determined by (i) the choice of
spreading kernel, (ii) the width of the
kernel (usually in number of grid points $w$), and (iii) the oversampling factor
$\sigma$~\cite{Dutt1993, Potts2003, Barnett2019}.
The computational complexity of a 3D NUFFT on $N$ non-uniform points can be
estimated to be $O(w^3 N + (\sigma M)^3 \log M)$.

\begin{figure}
  \begin{center}
    \includegraphics[]{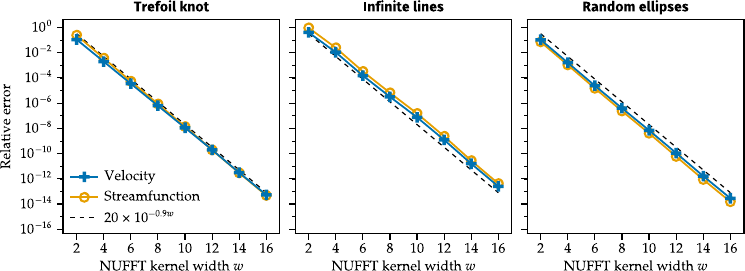}
  \end{center}
  \caption{%
    Velocity and streamfunction errors associated to the NUFFT precision,
    for the test cases illustrated in \cref{fig:visu_cases}.
    Relative root-mean-square error of long-range
    filament velocity $(+)$ and streamfunction values $(\times)$
    at varying NUFFT spreading widths $w$.
    NUFFTs are computed using the backwards Kaiser--Bessel spreading kernel
    \cref{eq:kaiser_bessel_bw} with an oversampling factor $\sigma = 1.5$.
    Dashed lines represent an exponential decay $\sim 10^{-0.9 w}$.
    In all cases, the long-range cut-off parameter is set to
    $\beta\longrange = 8$, ensuring subdominant cut-off errors.
    The Ewald splitting parameter is kept constant, $\alpha = 24 / L$.
  }\label{fig:accuracy_nufft}
\end{figure}

To evaluate NUFFTs, here we use the (backwards) Kaiser--Bessel function as spreading
kernel~\cite{Jackson1991, Potts2003}.
For a support in $x \in [-1, 1]$, the kernel and its Fourier transform are
\begin{equation}
  \phi(x) = \frac{\sinh \! {\left( \zeta \sqrt{1 - x^2} \right)}}{\pi \sqrt{1 - x^2}}
  \quad\leftrightarrow\quad
  \hat\phi(k) = I_0 \! \left( \sqrt{\zeta^2 - k^2} \right),
  \label{eq:kaiser_bessel_bw}
\end{equation}
where $I_0$ is the zero-th order modified Bessel function of the first kind.
The shape parameter is chosen as $\zeta = \gamma w \pi (1 - 1/2\sigma)$ to
minimise NUFFT errors~\cite{Potts2003}.
Here $\gamma = 0.995$ is an empirical \enquote{safety
factor}~\cite{Barnett2019} which slightly improves accuracy.
For performance reasons, the kernel $\phi(x)$ is evaluated via
a fast and accurate piecewise polynomial approximation~\cite{Barnett2019,
Shamshirgar2021} which takes advantage of SIMD vectorisation on modern CPUs.
Our NUFFT implementation~\cite{NonuniformFFTs} is written in Julia and is available on
\url{https://github.com/jipolanco/NonuniformFFTs.jl} (MIT License).
Compared to other existing libraries available in Julia~\cite{Barnett2019, Knopp2023},
our implementation allows real-valued non-uniform data, in which case it takes
advantage of real-to-complex FFT routines of the FFTW library~\cite{Frigo2005}.
Moreover, it uses the KernelAbstractions.jl~\cite{Churavy2024} package
to provide a platform-agnostic GPU implementation, enabling the acceleration of
vortex simulations using different kinds of GPUs.

In \cref{fig:accuracy_nufft}, we evaluate the effect of varying the NUFFT
accuracy on the relative error of long-range Biot--Savart computations,
for the cases illustrated in \cref{fig:visu_cases}.
Concretely, we set the oversampling factor to $\sigma = 1.5$ and vary the
spreading width $w$.
For comparison, we compute a reference solution with $\sigma = 2$ and
$w = 16$, which ensures that its accuracy is dominated by roundoff error.
As seen in \cref{fig:accuracy_nufft}, the relative error associated to the
long-range streamfunction and velocity decay exponentially as
$\sim 10^{-0.9w}$ when using an oversampling factor $\sigma = 1.5$.
For $\sigma = 2$, the same errors decay as $\sim 10^{-w}$ (not shown), but
at the expense of costlier FFTs.

\pagebreak  % force a page break to avoid having just the title of Appendix B at the end of the page

\section{Discretisation of spatial curves}\label{sm:sec:filament_discretisation}

The evaluation of the Biot--Savart law \cref{eq:BS} and its local
regularisation \cref{eq:BS_local} on vortex filaments require the estimation
of local tangent and curvature vectors from
a set of discrete points ${\left\{ \svec_j \right\}}_{j = 1}^N$ representing a
spatial curve.
This assumes that curves can be parametrised by some sufficiently regular
function $\svec(\tparam)$ whose derivatives on discrete locations $\tparam_j$ (such
that $\svec(\tparam_j) = \svec_j$) are well defined.
Furthermore, the use of quadratures to estimate line integrals (see \cref{eq:line_integral_quadrature})
requires being able to evaluate (interpolate) $\svec(\tparam)$ and its
derivatives at any arbitrary location $\tparam$.
To achieve these requirements, we choose to discretise curves using parametric
quintic splines.
First, note that in general the nodes $\svec_j$ are not equispaced, and that
the choice of the parametrisation (the $\tparam_j$ values) is arbitrary.
We adopt the common choice
of setting $\tparam_{j} = \tparam_{j - 1} + | \svec_{j} - \svec_{j - 1} |$ for
$j \in \left\{ 2, \ldots, N \right\}$, $\tparam_1 = 0$, where $| \cdot |$
represents the Euclidean distance.
In this case, $\tparam$ is a rough approximation of the curve arc length $\xi$,
tending to $\xi$ as the number of points increases.

We consider parametric splines of order $k$~\cite{Boor1978}.
Here by convention the polynomial order is $k - 1$, so that $k = 4$ and $6$
respectively correspond to cubic and quintic splines.
The basic idea is to parametrise a spatial curve in terms of a B-spline basis,
\begin{equation}
  \svec(\tparam) = \sum_{j = 1}^N \cvec_j b_j(\tparam),
  \quad\text{for } \tparam \in [0, \Tparam],
  \label{eq:spline_from_bspline_basis}
\end{equation}
where $b_j(\tparam)$ is a B-spline basis function.
The basis functions are fully defined by the choice of spline order $k$ and spline
\emph{knots} $t_j$, which are a set of locations in $[0, \Tparam]$.
Because the curves considered here are periodic, we can simply
set $t_j = \tparam_j$ for all $j$ (free-ended curves would require some extra care).
An important and convenient property of the B-splines is their compact support.
Namely, for even $k$, the B-spline $b_j(\tparam)$ is zero outside of the interval
$[t_{j - k/2}, t_{j + k/2}]$.

Spline interpolation consists in determining the coefficients $\cvec_j$ given
the equalities $\svec(\tparam_i) = \svec_i$ for $i \in \{ 1, \ldots, N \}$.
This leads to a linear system of the form $A_{ij} \cvec_j = \svec_i$ where
$A_{ij} = b_j(\tparam_i)$.
Thanks to the compact support of the B-splines, $\mat{A}$ is a cyclic banded matrix,
which enables the use of specialised algorithms to efficiently solve the system.
More precisely, with our choice of knots, $\mat{A}$ has $k - 1$ bands, as well
as a few out-of-bands entries in the top right and bottom left corners due to
periodicity.
In particular, for cubic splines ($k = 4$), $\mat{A}$ is almost tridiagonal
with two out-of-bands entries (one on each corner).
A standard solution is to convert the system to a fully tridiagonal one using
the Sherman--Morrison formula and then use the Thomas algorithm to solve the
system in $O(N)$ time~\cite{Yarrow1989}.
A similar procedure~\cite{Lv2008} can be applied to efficiently solve the cyclic
pentadiagonal system arising from periodic quintic spline interpolation ($k = 6$).

The derivative of a spline $\svec(\tparam)$ of order $k$ is itself a spline of
order $k - 1$, whose coefficients $\cvec'_j$ can be directly obtained from the
coefficients $\cvec_j$ using a local differentiation formula.
Moreover, splines are efficiently evaluated at locations $\tparam$ using de Boor's
algorithm~\cite{Boor1978}.
Therefore, knowing the coefficients $\cvec_j$ one can evaluate curve positions
and derivatives at any location $\tparam$.

\section{Derivation of truncation error estimates}\label{sm:sec:truncation_errors_derivation}

\subsection{Short-range errors}\label{sec:error_shortrange}

The absolute error associated to the truncation of the short-range velocity
integral
\cref{eq:velocity_shortrange} at a single location $\svec_0$ is given by
\begin{equation}
  \vvec\shortrange_{\text{err}}(\svec_0) =
  \frac{\kappa}{4\pi}
  \sum_{\nvec \in \mathbb{Z}^3}
  \int_{r > \rcut} g\shortrange(\rvec) \, \frac{\rvec \times \dd \svec}{|\rvec|^3}
  \quad
  \text{with } \rvec = \svec - \svec_0 + \nvec L,
\end{equation}
where the integral is over all vortex elements at a distance $|\rvec| = r > \rcut$
from the point $\svec_0$.
We first note that, for large $r$, the weight function
$g\shortrange(\rvec)$ defined in \cref{eq:ewald_derivative_weight_functions}
behaves as $g\shortrange(\rvec) = \frac{2\alpha r}{\sqrt{\pi}} e^{-{(\alpha r)}^2}
\left[ 1 + O(r^{-5/2}) \right]$.
Secondly, we assume that the vortices are homogeneously distributed in the
spatial domain, so that the total vortex length within a spherical shell of radii
$[r, r + \dd r]$ with infinitesimal thickness $\dd r$ is the product
between the mean vortex line density and the volume of the shell,
namely $\dd\VortexLength = (\VortexLength / V) (4\pi r^2 \dd r)$.
Here $\VortexLength$ is the total vortex length within a periodic cell of volume $V = L^3$.
Thirdly, we assume that the orientations of the vortex elements $\dd\svec$ within the
shell are random and independent of the separation vector $\rvec$.
Under these assumptions, we can formally estimate the root mean square (rms) short-range
velocity error to be
\begin{equation}
  \abserr\shortrange_v \approx
  \frac{\kappa}{4\pi}
  \int_{\rcut}^{\infty}
  \frac{2\alpha r}{\sqrt{\pi}} e^{-{(\alpha r)}^2} 
  \frac{r \, \dd\VortexLength}{r^3}
  =
  \frac{2 \kappa \VortexLength}{\sqrt{\pi} \alpha V}
  \int_{\beta\shortrange}^{\infty}
  x \, e^{-x^2} \, \dd x,
\end{equation}
where $\beta\shortrange = \alpha \rcut$.
This integrates exactly to
\begin{equation}
  \abserr\shortrange_v
  \approx \frac{\kappa \VortexLength}{\sqrt{\pi} \alpha V} \, e^{-{(\beta\shortrange)}^2}.
  \label{eq:error_shortrange_velocity}
\end{equation}
The same procedure can be applied to the short-range streamfunction \cref{eq:streamfunction_shortrange}.
Considering that
$\erfc(x) = \frac{e^{-x}}{x^{3/2}} \left[ 1 + O(x^{-2}) \right]$
for large $x$, one obtains the associated rms error
\begin{equation}
  \abserr\shortrange_\psi
  \approx
  \frac{\kappa}{4\pi}
  \int_{\rcut}^{\infty}
  \frac{e^{-{(\alpha r)}^2}}{{(\alpha r)}^{3/2}}
  \frac{\dd\VortexLength}{r}
  =
  \frac{\kappa \VortexLength}{\alpha^2 V}
  \int_{\beta\shortrange}^{\infty} \frac{e^{-x^2}}{\sqrt{x}} \, \dd x
  =
  \frac{\kappa \VortexLength}{2 \alpha^2 V}
  \, \Gamma \! \left( \frac{1}{4}, {\left(\beta\shortrange\right)}^2 \right).
  \label{eq:error_shortrange_streamfunction_with_gamma}
\end{equation}
Here $\Gamma(s, x) = \int_x^{\infty} u^{s - 1} e^{-u} \, \dd u$ is the
upper incomplete gamma function, which behaves asymptotically as
$\Gamma(s, x) = x^{s - 1} e^{-x} \left[ 1 + O(x^{-1}) \right]$.
Replacing this in \cref{eq:error_shortrange_streamfunction_with_gamma} finally
leads to the estimate
\begin{equation}
  \abserr\shortrange_\psi
  \approx \frac{\kappa \VortexLength}{2 \alpha^2 V}
  \, \frac{e^{-{(\beta\shortrange)}^2}}{{(\beta\shortrange)}^{3/2}}.
  \label{eq:error_shortrange_streamfunction}
\end{equation}
In both cases, the truncation error is dominated by the exponential decay
with the square of the non-dimensional cut-off parameter $\beta\sr$.

\subsection{Long-range errors}\label{sec:error_longrange}

The rms error associated to the long-range velocity field is defined by
\begin{equation}
  {\left(\abserr\longrange_{v}\right)}^2 =
  \frac{1}{V} \int_{\Omega} \left| {\vvec\longrange(\xvec) - \tilde{\vvec}\longrange(\xvec)} \right|^2 \, \dd^3\xvec,
\end{equation}
where $\vvec\longrange(\xvec)$ is the Fourier-truncated velocity field defined
in \cref{eq:velocity_longrange_physical}, and
$\tilde{\vvec}\longrange(\xvec)$ is the non-truncated velocity field
-- obtained by replacing $\mathbb{K}$ with
$\mathbb{K}_{\infty} \equiv \left\{ \frac{2\pi m}{L}; m \in \mathbb{Z} \right\}$ in
\cref{eq:velocity_longrange_physical}.
Using Parseval's theorem, the error can be equivalently written as
\begin{equation}
  {\left(\abserr\longrange_{v}\right)}^2 =
  \sum_{\kvec \in {(\mathbb{K}_{\infty} \backslash \, \mathbb{K})}^3}
  |\hat{\vvec}_{\kvec}\longrange|^2
  \;
  \le
  \;
  \sum_{|\kvec| > \kmax}
  |\hat{\vvec}_{\kvec}\longrange|^2,
  \label{eq:def_longrange_velocity_error_fourier}
\end{equation}
where $\hat{\vvec}_{\kvec}\longrange$ are the Fourier coefficients of the long-range velocity.
The last inequality results from converting from a \enquote{cubic} truncation
($k_i > \kmax$ for each Cartesian component $i$) to a spherical truncation $|\kvec| > \kmax$.

In physical terms, the velocity error \cref{eq:def_longrange_velocity_error_fourier}
can be directly related to the kinetic energy spectrum $E(k)$, defined in a
periodic domain as
\begin{equation}
  E(k) = \frac{1}{2 \Delta k} \sum_{|\kvec| \in \mathcal{I}_k}
  |\hat\vvec_{\kvec}|^2
  \quad\text{for } k = m \Delta k, \; m \in \mathbb{N},
  \label{eq:def_energy_spectrum}
\end{equation}
where $\hat\vvec_{\kvec}$ is a Fourier coefficient of the (unsmoothed) velocity field,
$\Delta k = 2\pi / L$ is the distance between two successive discrete
wavenumbers, and the sum is over all 3D wavenumbers $\kvec$ within the
spherical shell $\mathcal{I}_k \equiv \left[ k - \frac{\Delta k}{2}, k +
\frac{\Delta k}{2} \right)$.
Formally, the total kinetic energy (per unit mass) is then $E = \int_{0}^{\infty} E(k) \, \dd k$.
In the present context, \cref{eq:def_energy_spectrum} will allow us to quantify the effect of truncating
Fourier-space computations at a certain wavenumber $\kmax$.

At sufficiently small scales -- smaller than the typical
distance between vortices and the typical curvature radius of the
vortices -- vortex filaments can be considered as isolated straight lines.
Under that assumption,
one can show that the energy
spectrum is $E(k) = \frac{\kappa^2 \VortexLength}{4\pi L^3} \, k^{-1}$ for sufficiently
large $k$.  % ~\cite{Araki2002}.  % see footnote [14] in Araki et al.
To see this, one can consider the energy spectrum associated to a single straight vortex.
Without loss of generality, the vortex is aligned with the $z$ direction and passes through the origin, so
that its length within the cubic box is $\VortexLength = L$ and its
induced velocity field is invariant in $z$.
We can thus consider the simpler problem of a point vortex in
two-dimensional space.
By \cref{eq:vorticity_fourier_coefs}, the Fourier coefficients of its associated vorticity field are
$\hat\vortvec_{\kvec} = \frac{\kappa}{L^2} e^{-i \kvec \cdot \xvec} \, \vb{e}_z
= \frac{\kappa}{L^2} \vb{e}_z$,
where $\xvec = (x, y) = (0, 0)$ is the vortex location, $\kvec = (k_x, k_y)$, and
$\vb{e}_z$ is a unitary vector.
Then, the associated velocity coefficients are
$\hat\vvec_{\kvec} = i \kvec \times \frac{\hat\vortvec_{\kvec}}{k^2}$, whose squared magnitudes are
$|\hat\vvec_{\kvec}|^2 = \frac{\kappa^2}{L^4 k^2}$ (since $\hat\vortvec_{\kvec}$ is orthogonal to $\kvec$).
Besides, one can estimate the number of 2D Fourier modes intersecting the shell
$\mathcal{I}_k$ to be on average $N_k = 2 \pi m = k L$.
Since $|\hat\vvec_{\kvec}|^2$ only depends on the amplitude of $\kvec$,
one can assume that all non-zero $\hat\vvec_{\kvec}$ within the shell $\mathcal{I}_k$ are approximately equal.
Replacing in \cref{eq:def_energy_spectrum}, we finally obtain the energy spectrum
$E(k) \approx \frac{L}{4\pi} N_k |\hat\vvec_{\kvec}|^2
\approx \frac{\kappa^2}{4\pi L^2} k^{-1}
= \frac{\kappa^2 \VortexLength}{4\pi L^3} k^{-1}$.

\begin{figure}
  \begin{center}
    \includegraphics{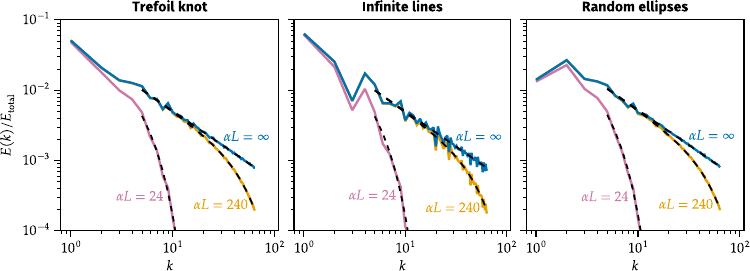}
  \end{center}
  \caption{%
    Kinetic energy spectra associated to three different test cases.
    The domain period is $L = 2\pi$ and the spectra are truncated at $\kmax = 64$
    (the number of Fourier modes in each direction is $M = 128$).
    Each panel shows the truncated energy spectrum associated to the unsmoothed
    velocity field $(\alpha L = \infty)$, as well as the energy spectra
    associated to two Gaussian-smoothed velocity fields ($\alpha L = 240 \text{ and } 24$).
    The spectra are normalised by the total kinetic energy $E_{\text{total}}$ obtained via
    \cref{eq:kinetic_energy_streamfunction}.
    At large wavenumbers $k$, the spectra are well approximated by
    $E(k) = \frac{\kappa^2 \VortexLength}{4\pi V} \, k^{-1} e^{-k^2 / 2\alpha^2}$
    (black dashed lines), where $\VortexLength$ is the total vortex length in
    each test case.
  }\label{fig:energy_spectra}
\end{figure}

This prediction is accurately verified in \cref{fig:energy_spectra} (blue curves) for the three
different vortex configurations illustrated in \cref{fig:visu_cases}.
Note that the relatively slow decay as $k^{-1}$ suggests that the integral of this
spectrum (and thus the total energy) presents a logarithmic divergence.
In fact, physically this is not the case, since the spectrum is regularised near the
wavenumber $k_a = 2\pi / \VortexRadius$ associated to the vortex core size
$\VortexRadius$, which is orders of magnitude smaller than the scales described by the VFM\@.

To provide an estimate of the error in the computation of the coarse-grained
velocity $\vvec\longrange$, we now consider its associated kinetic energy
spectrum $E\longrange(k) = E(k) \, e^{-k^2 / 2\alpha^2} = \frac{\kappa^2 \VortexLength}{4\pi V} \, k^{-1} e^{-k^2 / 2\alpha^2}$ for large $k$.
Note that the integral of $E\longrange(k)$ is the contribution
of the long-range velocity field to the total kinetic energy $E$.
Examples of $E\longrange(k)$ are shown in \cref{fig:energy_spectra} for two
different values of the smoothing parameter $\alpha$.
In the figure, the long-range truncation is at $\kmax = 64$ and the domain period is $L = 2\pi$.
It is clear that, if $\alpha$ is too large (e.g. $\alpha L = 240$ in the
figure, corresponding to $\beta\longrange = \kmax / 2\alpha \approx 0.84$), then this level of
truncation is insufficient to capture the whole
long-range kinetic energy.
A smaller value of $\alpha$ ($\alpha L = 24$ in the figure) means that the full
coarse-grained energy can be captured when truncating at the same $\kmax$.

Assuming the truncation wavenumber $\kmax$ to be sufficiently large, we can now
estimate an upper bound for the absolute truncation error
\cref{eq:def_longrange_velocity_error_fourier} associated to the long-range
velocity,
\begin{equation}
  \frac{{\left( \abserr\longrange_{v} \right)}^2}{2}
  \approx \int_{\kmax}^{\infty} E\longrange(k) \, \dd k
  = \frac{\kappa^2 \VortexLength}{4\pi V}
  \int_{\kmax}^{\infty} \frac{e^{-k^2 / 2\alpha^2}}{k} \, \dd k
  = \frac{\kappa^2 \VortexLength}{8\pi V} \,
  E_1 \! \left[2 {(\beta\longrange)}^2\right],
  \label{eq:error_longrange_energy}
\end{equation}
where $E_1(x) = \int_x^\infty \frac{e^{-u}}{u} \, \dd u$ is the exponential
integral function.
For large $x$, $E_1(x) = \frac{e^{-x}}{x} \left[ 1 - \frac{1}{x} + O(x^{-2}) \right]$,
and therefore
${\left( \abserr\longrange_{v} \right)}^2 \approx \frac{\kappa^2 \VortexLength}{8\pi V}
\, e^{-2 {(\beta\longrange)}^2} / {(\beta\longrange)}^2$.
Finally, the estimated error associated to the long-range velocity field is
\begin{equation}
  \abserr\longrange_v \approx
  \kappa \, {\left( \frac{\VortexLength}{8\pi V} \right)}^{\!\! 1/2} \,
  \frac{e^{-{(\beta\longrange)}^2}}{\beta\longrange}.
  \label{eq:error_longrange_velocity}
\end{equation}

A similar estimation can be done for the long-range truncation error
$\abserr\longrange_{\psi}$ associated to the streamfunction.
This is obtained by integrating
\begin{equation}
  \frac{{(\abserr\longrange_{\psi})}^2}{2}
  \approx \int_{\kmax}^{\infty} \frac{E\longrange(k)}{k^2} \, \dd k
  \approx \frac{\kappa^2}{4\alpha^2} \frac{\VortexLength}{4\pi V}
    \frac{e^{-2{(\beta\longrange)}^2}}{{[2{(\beta\longrange)}^2]}^2},
\end{equation}
where we have used
$\int_{x}^{\infty} \frac{e^{-u}}{u^2} \, \dd  u = \frac{e^{-x}}{x} - E_1(x) = \frac{e^{-x}}{x^2} \left[ 1 + O(x^{-1}) \right]$.
This results in
\begin{equation}
  \abserr\longrange_\psi \approx
  \frac{\kappa}{2\alpha} \,
  {\left( \frac{\VortexLength}{8\pi V} \right)}^{\!\! 1/2} \,
  \frac{e^{-{(\beta\longrange)}^2}}{{(\beta\longrange)}^2}.
  \label{eq:error_longrange_streamfunction}
\end{equation}

Similarly to short-range errors, the long-range truncation errors are dominated
by the exponential decay with the square of the cut-off parameter $\beta\lr$.
We stress that the estimates
\cref{eq:error_longrange_velocity,eq:error_longrange_streamfunction} are
expected to be valid for sufficiently large values of $\kmax$ (or
$\beta\longrange$).
Furthermore, they can be expected to overestimate the actual errors due to the
inequality in \cref{eq:def_longrange_velocity_error_fourier}, which discards
the contributions of some large-wavenumber modes which are actually resolved in
the simulations (more precisely, a factor of $\sim \! 6/\pi$ Fourier modes is
discarded by the above estimates).

\subsection{Combined truncation errors}\label{sec:error_combined}

To avoid unnecessary computations and achieve optimal accuracy for a given
computational cost, one would like the errors associated to the short- and long-range
components to be equivalent.
In light of the above estimations, it seems natural to set their respective
non-dimensional cut-off parameters to the same value, $\beta\sr = \beta\lr = \beta$, which also
helps reducing the size of the parameter space.
This finally leads to the following total truncation error estimates:
\begin{align}
  \abserr_v
  &\approx
  \kappa \left[
    \frac{\VortexLength}{\sqrt{\pi} \alpha V}
    +
    \frac{1}{\beta} {\left( \frac{\VortexLength}{8\pi V} \right)}^{\!\! 1/2}
  \right]
  e^{-\beta^2},
  \\
  \abserr_\psi
  &\approx
  \frac{\kappa}{2\alpha}
  \left[
    \frac{\VortexLength}{\alpha V}
    +
    \frac{1}{\beta^{1/2}}
    {\left( \frac{\VortexLength}{8\pi V} \right)}^{\!\! 1/2}
  \right]
  \frac{e^{-\beta^2}}{\beta^{3/2}}.
\end{align}
In summary, the accuracy of the method is mainly controlled by a unique
non-dimensional parameter $\beta$, from which both physical- and Fourier-space
cut-offs $\rcut$ and $\kmax$ can be obtained.
This still leaves the inverse splitting distance $\alpha$ as a free parameter
that can be adjusted to optimise performance (\cref{sec:numerical:performance}).
  % arXiv: display supplementary material as extra appendices

\bibliographystyle{siamplain}
\bibliography{VFM}

\begin{thebibliography}{10}

\bibitem{Ambrose2013}
{\sc D.~M. Ambrose, M.~Siegel, and S.~Tlupova}, {\em A small-scale
  decomposition for {{3D}} boundary integral computations with surface
  tension}, J. Comput. Phys., 247 (2013), pp.~168--191,
  \url{https://doi.org/10.1016/j.jcp.2013.03.045}.

\bibitem{Arms1965}
{\sc R.~J. Arms and F.~R. Hama}, {\em Localized-{{Induction Concept}} on a
  {{Curved Vortex}} and {{Motion}} of an {{Elliptic Vortex Ring}}}, Phys.
  Fluids, 8 (1965), pp.~553--559, \url{https://doi.org/10.1063/1.1761268}.

\bibitem{Arnold2013}
{\sc A.~Arnold, F.~Fahrenberger, C.~Holm, O.~Lenz, M.~Bolten, H.~Dachsel,
  R.~Halver, I.~Kabadshow, F.~G{\"a}hler, F.~Heber, J.~Iseringhausen,
  M.~Hofmann, M.~Pippig, D.~Potts, and G.~Sutmann}, {\em Comparison of scalable
  fast methods for long-range interactions}, Phys. Rev. E, 88 (2013),
  p.~063308, \url{https://doi.org/10.1103/PhysRevE.88.063308}.

\bibitem{Arnold2005}
{\sc A.~Arnold and C.~Holm}, {\em Efficient {{Methods}} to {{Compute Long-Range
  Interactions}} for {{Soft Matter Systems}}}, in Advanced {{Computer
  Simulation Approaches}} for {{Soft Matter Sciences II}}, C.~Holm and
  K.~Kremer, eds., Advances in {{Polymer Science}}, Springer, Berlin,
  Heidelberg, 2005, pp.~59--109, \url{https://doi.org/10.1007/b136793}.

\bibitem{Baggaley2011}
{\sc A.~W. Baggaley and C.~F. Barenghi}, {\em Spectrum of turbulent
  {{Kelvin-waves}} cascade in superfluid helium}, Phys. Rev. B, 83 (2011),
  p.~134509, \url{https://doi.org/10.1103/PhysRevB.83.134509}.

\bibitem{Baggaley2012f}
{\sc A.~W. Baggaley and C.~F. Barenghi}, {\em Tree {{Method}} for {{Quantum
  Vortex Dynamics}}}, J. Low Temp. Phys., 166 (2012), pp.~3--20,
  \url{https://doi.org/10.1007/s10909-011-0405-6}.

\bibitem{Baggaley2014}
{\sc A.~W. Baggaley and J.~Laurie}, {\em Kelvin-wave cascade in the vortex
  filament model}, Phys. Rev. B, 89 (2014), p.~014504,
  \url{https://doi.org/10.1103/PhysRevB.89.014504}.

\bibitem{Baggaley2012}
{\sc A.~W. Baggaley, J.~Laurie, and C.~F. Barenghi}, {\em Vortex-{{Density
  Fluctuations}}, {{Energy Spectra}}, and {{Vortical Regions}} in {{Superfluid
  Turbulence}}}, Phys. Rev. Lett., 109 (2012), p.~205304,
  \url{https://doi.org/10.1103/PhysRevLett.109.205304}.

\bibitem{Ballenegger2008}
{\sc V.~Ballenegger, J.~J. Cerda, O.~Lenz, and {\relax Ch}.~Holm}, {\em The
  optimal {{P3M}} algorithm for computing electrostatic energies in periodic
  systems}, J. Chem. Phys., 128 (2008), p.~034109,
  \url{https://doi.org/10.1063/1.2816570}.

\bibitem{Banica2024}
{\sc V.~Banica, G.~Maierhofer, and K.~Schratz}, {\em Numerical {{Integration}}
  of {{Schr{\"o}dinger Maps}} via the {{Hasimoto Transform}}}, SIAM J. Numer.
  Anal., 62 (2024), pp.~322--352, \url{https://doi.org/10.1137/22M1531555}.

\bibitem{Barenghi2014}
{\sc C.~F. Barenghi, V.~S. L'vov, and P.-E. Roche}, {\em Experimental,
  numerical, and analytical velocity spectra in turbulent quantum fluid}, Proc.
  Natl. Acad. Sci. USA, 111 (2014), pp.~4683--4690,
  \url{https://doi.org/10.1073/pnas.1312548111}.

\bibitem{Barenghi2014a}
{\sc C.~F. Barenghi, L.~Skrbek, and K.~R. Sreenivasan}, {\em Introduction to
  quantum turbulence}, Proc. Natl. Acad. Sci. USA, 111 (2014), pp.~4647--4652,
  \url{https://doi.org/10.1073/pnas.1400033111}.

\bibitem{Barenghi2023}
{\sc C.~F. Barenghi, L.~Skrbek, and K.~R. Sreenivasan}, {\em Quantum
  {{Turbulence}}}, Cambridge University Press, Cambridge, 2023,
  \url{https://doi.org/10.1017/9781009345651}.

\bibitem{Barnes1986}
{\sc J.~Barnes and P.~Hut}, {\em A hierarchical {{O}}({{N}} log {{N}})
  force-calculation algorithm}, Nature, 324 (1986), pp.~446--449,
  \url{https://doi.org/10.1038/324446a0}.

\bibitem{Barnett2019}
{\sc A.~H. Barnett, J.~Magland, and L.~{af Klinteberg}}, {\em A {{Parallel
  Nonuniform Fast Fourier Transform Library Based}} on an ``{{Exponential}} of
  {{Semicircle}}" {{Kernel}}}, SIAM J. Sci. Comput., 41 (2019), pp.~C479--C504,
  \url{https://doi.org/10.1137/18M120885X}.

\bibitem{Batchelor1970}
{\sc G.~K. Batchelor}, {\em Slender-body theory for particles of arbitrary
  cross-section in {{Stokes}} flow}, J. Fluid Mech., 44 (1970), pp.~419--440,
  \url{https://doi.org/10.1017/S002211207000191X}.

\bibitem{Bezanson2017}
{\sc J.~Bezanson, A.~Edelman, S.~Karpinski, and V.~B. Shah}, {\em Julia: {{A}}
  fresh approach to numerical computing}, SIAM Rev., 59 (2017), pp.~65--98,
  \url{https://doi.org/10.1137/141000671}.

\bibitem{Blanes2024}
{\sc S.~Blanes, F.~Casas, and A.~Murua}, {\em Splitting {{Methods}} for
  differential equations}, May 2024, \url{https://arxiv.org/abs/2401.01722}.
\newblock To appear in Acta Numer. (2024).

\bibitem{Boyd2001}
{\sc J.~P. Boyd}, {\em Chebyshev and {{Fourier Spectral Methods}}}, Dover
  Publications, Mineola, N.Y, second edition~ed., Dec. 2001.

\bibitem{Buttke1988}
{\sc T.~F. Buttke}, {\em A numerical study of superfluid turbulence in the
  self-induction approximation}, Journal of Computational Physics, 76 (1988),
  pp.~301--326, \url{https://doi.org/10.1016/0021-9991(88)90145-3}.

\bibitem{Callegari1978}
{\sc A.~J. Callegari and L.~Ting}, {\em Motion of a {{Curved Vortex Filament}}
  with {{Decaying Vortical Core}} and {{Axial Velocity}}}, SIAM J. Appl. Math.,
  35 (1978), pp.~148--175, \url{https://doi.org/10.1137/0135013}.

\bibitem{Canuto1988}
{\sc C.~Canuto, M.~Y. Hussaini, A.~Quarteroni, and T.~A. Zang}, {\em Spectral
  {{Methods}} in {{Fluid Dynamics}}}, Springer Berlin Heidelberg, Berlin,
  Heidelberg, 1988, \url{https://doi.org/10.1007/978-3-642-84108-8}.

\bibitem{Churavy2024}
{\sc V.~Churavy}, {\em {KernelAbstractions}.jl}.
\newblock Zenodo, Dec. 2024, \url{https://doi.org/10.5281/zenodo.4021259}.

\bibitem{Cichocki1989}
{\sc B.~Cichocki and B.~U. Felderhof}, {\em Electrostatic interactions in
  two-dimensional {{Coulomb}} systems with periodic boundary conditions},
  Physica A, 158 (1989), pp.~706--722,
  \url{https://doi.org/10.1016/0378-4371(89)90487-1}.

\bibitem{Cottet2000}
{\sc G.-H. Cottet and P.~D. Koumoutsakos}, {\em Vortex {{Methods}}: {{Theory}}
  and {{Practice}}}, Cambridge University Press, Cambridge, 2000,
  \url{https://doi.org/10.1017/CBO9780511526442}.

\bibitem{Darden1993}
{\sc T.~Darden, D.~York, and L.~Pedersen}, {\em Particle mesh {{Ewald}}: {{An
  N}}{$\cdot$}log({{N}}) method for {{Ewald}} sums in large systems}, J. Chem.
  Phys., 98 (1993), pp.~10089--10092, \url{https://doi.org/10.1063/1.464397}.

\bibitem{Boor1978}
{\sc C.~de~Boor}, {\em A {{Practical Guide}} to {{Splines}}}, Applied
  {{Mathematical Sciences}}, Springer-Verlag, New York, 1978.

\bibitem{DeLaHoz2009}
{\sc F.~De~La~Hoz, C.~J. {Garc{\'i}a-Cervera}, and L.~Vega}, {\em A {{Numerical
  Study}} of the {{Self-Similar Solutions}} of the {{Schr{\"o}dinger Map}}},
  SIAM J. Appl. Math., 70 (2009), pp.~1047--1077,
  \url{https://arxiv.org/abs/27862547}.

\bibitem{Deserno1998}
{\sc M.~Deserno and C.~Holm}, {\em How to mesh up {{Ewald}} sums. {{I}}. {{A}}
  theoretical and numerical comparison of various particle mesh routines}, J.
  Chem. Phys., 109 (1998), pp.~7678--7693,
  \url{https://doi.org/10.1063/1.477414}.

\bibitem{Deserno1998a}
{\sc M.~Deserno and C.~Holm}, {\em How to mesh up {{Ewald}} sums. {{II}}.
  {{An}} accurate error estimate for the particle--particle--particle-mesh
  algorithm}, J. Chem. Phys., 109 (1998), pp.~7694--7701,
  \url{https://doi.org/10.1063/1.477415}.

\bibitem{Dutt1993}
{\sc A.~Dutt and V.~Rokhlin}, {\em Fast {{Fourier Transforms}} for
  {{Nonequispaced Data}}}, SIAM J. Sci. Comput., 14 (1993), pp.~1368--1393,
  \url{https://doi.org/10.1137/0914081}.

\bibitem{Ewald1921}
{\sc P.~P. Ewald}, {\em Die {{Berechnung}} optischer und elektrostatischer
  {{Gitterpotentiale}}}, Ann. Phys., 369 (1921), pp.~253--287,
  \url{https://doi.org/10.1002/andp.19213690304}.

\bibitem{Feynman2011}
{\sc R.~P. Feynman, R.~B. Leighton, and M.~L. Sands}, {\em The {{Feynman}}
  Lectures on Physics. {{Volume}} 2: {{Mainly}} Electromagnetism and Matter},
  Basic Books, New York, 2011.

\bibitem{Frigo2005}
{\sc M.~Frigo and S.~G. Johnson}, {\em The design and implementation of
  {{FFTW3}}}, Proc. IEEE, 93 (2005), pp.~216--231,
  \url{https://doi.org/10.1109/JPROC.2004.840301}.

\bibitem{Galantucci2020}
{\sc L.~Galantucci, A.~W. Baggaley, C.~F. Barenghi, and G.~Krstulovic}, {\em A
  new self-consistent approach of quantum turbulence in superfluid helium},
  Eur. Phys. J. Plus, 135 (2020), p.~547,
  \url{https://doi.org/10.1140/epjp/s13360-020-00543-0}.

\bibitem{Gazzola2014}
{\sc M.~Gazzola, B.~Hejazialhosseini, and P.~Koumoutsakos}, {\em Reinforcement
  {{Learning}} and {{Wavelet Adapted Vortex Methods}} for {{Simulations}} of
  {{Self-propelled Swimmers}}}, SIAM J. Sci. Comput., 36 (2014),
  pp.~B622--B639, \url{https://doi.org/10.1137/130943078}.

\bibitem{Greengard2004}
{\sc L.~Greengard and J.-Y. Lee}, {\em Accelerating the {{Nonuniform Fast
  Fourier Transform}}}, SIAM Rev., 46 (2004), pp.~443--454,
  \url{https://doi.org/10.1137/S003614450343200X}.

\bibitem{Greengard1987}
{\sc L.~Greengard and V.~Rokhlin}, {\em A fast algorithm for particle
  simulations}, J. Comput. Phys., 73 (1987), pp.~325--348,
  \url{https://doi.org/10.1016/0021-9991(87)90140-9}.

\bibitem{Grotto2024}
{\sc F.~Grotto and S.~Morlacchi}, {\em Decay of time correlations in point
  vortex systems}, Physica D,  (2024), p.~134169,
  \url{https://doi.org/10.1016/j.physd.2024.134169}.

\bibitem{Hanninen2013}
{\sc R.~H{\"a}nninen}, {\em Dissipation enhancement from a single vortex
  reconnection in superfluid helium}, Phys. Rev. B, 88 (2013), p.~054511,
  \url{https://doi.org/10.1103/PhysRevB.88.054511}.

\bibitem{Hanninen2014}
{\sc R.~H{\"a}nninen and A.~W. Baggaley}, {\em Vortex filament method as a tool
  for computational visualization of quantum turbulence}, Proc. Natl. Acad.
  Sci. USA, 111 (2014), pp.~4667--4674,
  \url{https://doi.org/10.1073/pnas.1312535111}.

\bibitem{Hasimoto1959}
{\sc H.~Hasimoto}, {\em On the periodic fundamental solutions of the {{Stokes}}
  equations and their application to viscous flow past a cubic array of
  spheres}, J. Fluid Mech., 5 (1959), pp.~317--328,
  \url{https://doi.org/10.1017/S0022112059000222}.

\bibitem{Hasimoto1972}
{\sc H.~Hasimoto}, {\em A soliton on a vortex filament}, J. Fluid Mech., 51
  (1972), pp.~477--485, \url{https://doi.org/10.1017/S0022112072002307}.

\bibitem{Hedman2006}
{\sc F.~Hedman and A.~Laaksonen}, {\em Ewald summation based on nonuniform fast
  {{Fourier}} transform}, Chem. Phys. Lett., 425 (2006), pp.~142--147,
  \url{https://doi.org/10.1016/j.cplett.2006.04.106}.

\bibitem{Helmholtz1858}
{\sc H.~Helmholtz}, {\em {{\"U}ber Integrale der hydrodynamischen Gleichungen,
  welche den Wirbelbewegungen entsprechen.}}, J. Reine Angew. Math., 55 (1858),
  pp.~25--55.

\bibitem{Hockney1988}
{\sc R.~Hockney and J.~Eastwood}, {\em Computer {{Simulation Using
  Particles}}}, CRC Press, 1988, \url{https://doi.org/10.1201/9780367806934}.

\bibitem{Jackson1991}
{\sc J.~Jackson, C.~Meyer, D.~Nishimura, and A.~Macovski}, {\em Selection of a
  convolution function for {{Fourier}} inversion using gridding (computerised
  tomography application)}, IEEE Trans. Med. Imaging, 10 (Sept./1991),
  pp.~473--478, \url{https://doi.org/10.1109/42.97598}.

\bibitem{Kivotides2000}
{\sc D.~Kivotides, C.~F. Barenghi, and D.~C. Samuels}, {\em Triple {{Vortex
  Ring Structure}} in {{Superfluid Helium II}}}, Science, 290 (2000),
  pp.~777--779, \url{https://doi.org/10.1126/science.290.5492.777}.

\bibitem{Knopp2023}
{\sc T.~Knopp, M.~Boberg, and M.~Grosser}, {\em {{NFFT}}.jl: {{Generic}} and
  {{Fast Julia Implementation}} of the {{Nonequidistant Fast Fourier
  Transform}}}, SIAM J. Sci. Comput., 45 (2023), pp.~C179--C205,
  \url{https://doi.org/10.1137/22M1510935}.

\bibitem{Kolafa1992}
{\sc J.~Kolafa and J.~W. Perram}, {\em Cutoff {{Errors}} in the {{Ewald
  Summation Formulae}} for {{Point Charge Systems}}}, Mol. Simul., 9 (1992),
  pp.~351--368, \url{https://doi.org/10.1080/08927029208049126}.

\bibitem{Koumoutsakos2005}
{\sc P.~Koumoutsakos}, {\em Multiscale {{Flow Simulations Using Particles}}},
  Annu. Rev. Fluid Mech., 37 (2005), pp.~457--487,
  \url{https://doi.org/10.1146/annurev.fluid.37.061903.175753}.

\bibitem{Krishnamurthy2023}
{\sc V.~S. Krishnamurthy and T.~Sakajo}, {\em The {{N-vortex}} problem in a
  doubly periodic rectangular domain with constant background vorticity},
  Physica D, 448 (2023), p.~133728,
  \url{https://doi.org/10.1016/j.physd.2023.133728}.

\bibitem{Lamb1945}
{\sc H.~Lamb}, {\em Hydrodynamics}, Dover Publications, 6~ed., 1945,
  \url{https://archive.org/details/hydrodynamics00lamb}.

\bibitem{Lauga2009}
{\sc E.~Lauga and T.~R. Powers}, {\em The hydrodynamics of swimming
  microorganisms}, Rep. Prog. Phys., 72 (2009), p.~096601,
  \url{https://doi.org/10.1088/0034-4885/72/9/096601}.

\bibitem{Laurie2023}
{\sc J.~Laurie and A.~W. Baggaley}, {\em Vorticity locking and pressure
  dynamics in finite-temperature superfluid turbulence}, Phys. Rev. Fluids, 8
  (2023), p.~054604, \url{https://doi.org/10.1103/PhysRevFluids.8.054604}.

\bibitem{Lindbo2010}
{\sc D.~Lindbo and A.-K. Tornberg}, {\em Spectrally accurate fast summation for
  periodic {{Stokes}} potentials}, J. Comput. Phys., 229 (2010),
  pp.~8994--9010, \url{https://doi.org/10.1016/j.jcp.2010.08.026}.

\bibitem{Lv2008}
{\sc X.-G. Lv and J.~Le}, {\em A note on solving nearly penta-diagonal linear
  systems}, Appl. Math. Comput., 204 (2008), pp.~707--712,
  \url{https://doi.org/10.1016/j.amc.2008.07.012}.

\bibitem{Mattson1999}
{\sc W.~Mattson and B.~M. Rice}, {\em Near-neighbor calculations using a
  modified cell-linked list method}, Comput. Phys. Commun., 119 (1999),
  pp.~135--148, \url{https://doi.org/10.1016/S0010-4655(98)00203-3}.

\bibitem{Maxian2021}
{\sc O.~Maxian, A.~Mogilner, and A.~Donev}, {\em Integral-based spectral method
  for inextensible slender fibers in {{Stokes}} flow}, Phys. Rev. Fluids, 6
  (2021), p.~014102, \url{https://doi.org/10.1103/PhysRevFluids.6.014102}.

\bibitem{McLachlan2002}
{\sc R.~I. McLachlan and G.~R.~W. Quispel}, {\em Splitting methods}, Acta
  Numer., 11 (2002), pp.~341--434,
  \url{https://doi.org/10.1017/S0962492902000053}.

\bibitem{Monaghan1993}
{\sc J.~J. Monaghan and R.~J. Humble}, {\em Vortex {{Particle Methods}} for
  {{Periodic Channel Flow}}}, J. Comput. Phys., 107 (1993), pp.~152--159,
  \url{https://doi.org/10.1006/jcph.1993.1132}.

\bibitem{Moore1972}
{\sc D.~W. Moore and P.~G. Saffman}, {\em The motion of a vortex filament with
  axial flow}, Phil. Trans. R. Soc. Lond. A, 272 (1972), pp.~403--429,
  \url{https://doi.org/10.1098/rsta.1972.0055}.

\bibitem{Mori2020}
{\sc Y.~Mori, L.~Ohm, and D.~Spirn}, {\em Theoretical {{Justification}} and
  {{Error Analysis}} for {{Slender Body Theory}}}, Commun. Pure Appl. Math., 73
  (2020), pp.~1245--1314, \url{https://doi.org/10.1002/cpa.21872}.

\bibitem{Muller2021}
{\sc N.~P. M{\"u}ller, J.~I. Polanco, and G.~Krstulovic}, {\em Intermittency of
  {{Velocity Circulation}} in {{Quantum Turbulence}}}, Phys. Rev. X, 11 (2021),
  p.~011053, \url{https://doi.org/10.1103/PhysRevX.11.011053}.

\bibitem{Newton2009}
{\sc P.~K. Newton and G.~Chamoun}, {\em Vortex {{Lattice Theory}}: {{A Particle
  Interaction Perspective}}}, SIAM Rev., 51 (2009), pp.~501--542,
  \url{https://doi.org/10.1137/07068597x}.

\bibitem{Pippig2013}
{\sc M.~Pippig and D.~Potts}, {\em Parallel {{Three-Dimensional Nonequispaced
  Fast Fourier Transforms}} and {{Their Application}} to {{Particle
  Simulation}}}, SIAM J. Sci. Comput., 35 (2013), pp.~C411--C437,
  \url{https://doi.org/10.1137/120888478}.

\bibitem{NonuniformFFTs}
{\sc J.~I. Polanco}, {\em {{NonuniformFFTs}}.jl}.
\newblock Zenodo, Nov. 2024, \url{https://doi.org/10.5281/zenodo.14637607}.

\bibitem{VortexPasta}
{\sc J.~I. Polanco}, {\em {{VortexPasta}}.jl}.
\newblock Zenodo, Jan. 2025, \url{https://doi.org/10.5281/zenodo.14749388}.

\bibitem{Polanco2021}
{\sc J.~I. Polanco, N.~P. M{\"u}ller, and G.~Krstulovic}, {\em Vortex
  clustering, polarisation and circulation intermittency in classical and
  quantum turbulence}, Nat. Commun., 12 (2021), p.~7090,
  \url{https://doi.org/10.1038/s41467-021-27382-6}.

\bibitem{Potts2003}
{\sc D.~Potts and G.~Steidl}, {\em Fast {{Summation}} at {{Nonequispaced
  Knots}} by {{NFFTs}}}, SIAM J. Sci. Comput., 24 (2003), pp.~2013--2037,
  \url{https://doi.org/10.1137/S1064827502400984}.

\bibitem{Roberts1970}
{\sc P.~H. Roberts and R.~J. Donnelly}, {\em Dynamics of vortex rings}, Phys.
  Lett. A, 31 (1970), pp.~137--138,
  \url{https://doi.org/10.1016/0375-9601(70)90193-3}.

\bibitem{Saffman1993}
{\sc P.~G. Saffman}, {\em Vortex {{Dynamics}}}, Cambridge University Press,
  Jan. 1993, \url{https://doi.org/10.1017/cbo9780511624063}.

\bibitem{Saintillan2005}
{\sc D.~Saintillan, E.~Darve, and E.~S.~G. Shaqfeh}, {\em A smooth
  particle-mesh {{Ewald}} algorithm for {{Stokes}} suspension simulations:
  {{The}} sedimentation of fibers}, Phys. Fluids, 17 (2005), p.~033301,
  \url{https://doi.org/10.1063/1.1862262}.

\bibitem{Samuels1992}
{\sc D.~C. Samuels}, {\em Velocity matching and {{Poiseuille}} pipe flow of
  superfluid helium}, Phys. Rev. B, 46 (1992), pp.~11714--11724,
  \url{https://doi.org/10.1103/PhysRevB.46.11714}.

\bibitem{Samuels2001}
{\sc D.~C. Samuels}, {\em Vortex {{Filament Methods}} for {{Superfluids}}}, in
  Quantized {{Vortex Dynamics}} and {{Superfluid Turbulence}}, C.~F. Barenghi,
  R.~J. Donnelly, and W.~F. Vinen, eds., Lecture {{Notes}} in {{Physics}},
  Springer, Berlin, Heidelberg, 2001, pp.~97--113,
  \url{https://doi.org/10.1007/3-540-45542-6_9}.

\bibitem{Sandu2019}
{\sc A.~Sandu}, {\em A {{Class}} of {{Multirate Infinitesimal GARK Methods}}},
  SIAM J. Numer. Anal., 57 (2019), pp.~2300--2327,
  \url{https://doi.org/10.1137/18m1205492}.

\bibitem{Schwarz1985}
{\sc K.~W. Schwarz}, {\em Three-dimensional vortex dynamics in superfluid
  {\textsuperscript{4}}{{He}}: {{Line-line}} and line-boundary interactions},
  Phys. Rev. B, 31 (1985), pp.~5782--5804,
  \url{https://doi.org/10.1103/PhysRevB.31.5782}.

\bibitem{Shamshirgar2021}
{\sc D.~S. Shamshirgar, J.~Bagge, and A.-K. Tornberg}, {\em Fast {{Ewald}}
  summation for electrostatic potentials with arbitrary periodicity}, J. Chem.
  Phys., 154 (2021), p.~164109, \url{https://doi.org/10.1063/5.0044895}.

\bibitem{Shariff1992}
{\sc K.~Shariff and A.~Leonard}, {\em Vortex {{Rings}}}, Annu. Rev. Fluid
  Mech., 24 (1992), pp.~235--279,
  \url{https://doi.org/10.1146/annurev.fl.24.010192.001315}.

\bibitem{Sullivan2008}
{\sc I.~S. Sullivan, J.~J. Niemela, R.~E. Hershberger, D.~Bolster, and R.~J.
  Donnelly}, {\em Dynamics of thin vortex rings}, J. Fluid Mech., 609 (2008),
  pp.~319--347, \url{https://doi.org/10.1017/S0022112008002292}.

\bibitem{Thomson1880}
{\sc W.~Thomson}, {\em 3. {{Vibrations}} of a {{Columnar Vortex}}}, Proc. R.
  Soc. Edinb., 10 (1880/ed), pp.~443--456,
  \url{https://doi.org/10.1017/S0370164600044151}.

\bibitem{vanKan2021a}
{\sc A.~{van Kan}, A.~Alexakis, and M.-E. Brachet}, {\em Intermittency of
  three-dimensional perturbations in a point-vortex model}, Phys. Rev. E, 103
  (2021), p.~053102, \url{https://doi.org/10.1103/PhysRevE.103.053102}.

\bibitem{Vermeer2003}
{\sc L.~J. Vermeer, J.~N. S{\o}rensen, and A.~Crespo}, {\em Wind turbine wake
  aerodynamics}, Prog. Aerosp. Sci., 39 (2003), pp.~467--510,
  \url{https://doi.org/10.1016/S0376-0421(03)00078-2}.

\bibitem{Wacks2014}
{\sc D.~H. Wacks, A.~W. Baggaley, and C.~F. Barenghi}, {\em Coherent laminar
  and turbulent motion of toroidal vortex bundles}, Phys. Fluids, 26 (2014),
  p.~027102, \url{https://doi.org/10.1063/1.4864659}.

\bibitem{Weiss1991}
{\sc J.~B. Weiss and J.~C. McWilliams}, {\em Nonergodicity of point vortices},
  Phys. Fluids, 3 (1991), pp.~835--844, \url{https://doi.org/10.1063/1.858014}.

\bibitem{Yarrow1989}
{\sc M.~Yarrow}, {\em Solving periodic block tridiagonal systems using the
  {{Sherman-Morrison-Woodbury}} formula}, in 9th {{Comput}}. {{Fluid Dyn}}.
  {{Conf}}., Buffalo, NY, U.S.A., June 1989, {American Institute of Aeronautics
  and Astronautics}, \url{https://doi.org/10.2514/6.1989-1946}.

\bibitem{Yui2021}
{\sc S.~Yui, H.~Kobayashi, M.~Tsubota, and R.~Yokota}, {\em Quantum turbulence
  coupled with externally driven normal-fluid turbulence in superfluid
  {\textsuperscript{4}}{{He}}}, May 2021,
  \url{https://doi.org/10.48550/arXiv.2105.09499},
  \url{https://arxiv.org/abs/2105.09499}.

\bibitem{Yui2018}
{\sc S.~Yui, M.~Tsubota, and H.~Kobayashi}, {\em Three-{{Dimensional Coupled
  Dynamics}} of the {{Two-Fluid Model}} in {{Superfluid}}
  {\textsuperscript{4}}{{He}}: {{Deformed Velocity Profile}} of {{Normal
  Fluid}} in {{Thermal Counterflow}}}, Phys. Rev. Lett., 120 (2018), p.~155301,
  \url{https://doi.org/10.1103/PhysRevLett.120.155301}.

\end{thebibliography}

\end{document}